\documentclass[intlimits,twoside,a4paper]{article}

\usepackage{amsmath,amssymb}
\usepackage{graphicx}
\usepackage{wrapfig}

\usepackage[T2A]{fontenc}
\usepackage[cp1251]{inputenc}
\usepackage[eqsecnum]{cmpj}

\issue{2011}{14}{2}{23603}
\doinumber{10.5488/CMP.14.23603}

\title[Phase transitions in the Mitsui model]%
{Phase transitions in the Mitsui model%
}
\author[Yu.I. Dublenych]{Yu.I. Dublenych}
\address{Institute for Condensed Matter Physics of the National
Academy of Sciences of Ukraine,  \\ 1~Svientsitskii Str.,
79011~Lviv, Ukraine }

\authorcopyright{Yu.I. Dublenych, 2011}

\date{Received April 14, 2011, in final form June 14, 2011}

\begin{document}

\maketitle

\begin{abstract}
In this paper, phase transitions in the Mitsui model without
longitudinal field but with a transverse one are investigated in
the mean field approximation. The one-to-one correspondence has
been established between this model and the two-sublattice
Ising-type model with longitudinal and transverse fields. Phase
diagrams and diagrams of existence of the ferroelectric phase are
constructed. In the case $\Omega =0$ ($\Omega$ is the transverse
field), a simple analytical expression for the tricritical
temperature and the condition of existence of the tricritical
point are obtained. For $\Omega \ne 0$, systems of equations for the
tricritical point and for the condition of its existence are
written.
\pacs 64.60.De, 64.60.Cn, 77.80.-e
\end{abstract}

\section{Introduction}
The Mitsui model was proposed in 1958~\cite{bib1} to
theoretically explain the ferroelectric properties of the Rochelle salt.
In 1971, \v Zek\v s, Shukla, and Blinc~\cite{bib2} formulated this
model in terms of pseudospins and just in this form  it is known
at the present time.

For better quantitative description of the Rochelle salt the
conventional Mitsui model was modified in many ways. For instance,
in reference~\cite{bib3} an additional piezoelectric interaction
was included in the Hamiltonian of the model and in
references~\cite{bib4} and~\cite{bib5} a transverse field (or
tunneling) was included as well. To take into account a realistic
structure of Rochelle salt crystal, the four-sublattice Mitsui
model was considered~\cite{bib6}. Besides the Rochelle salt, the
Mitsui model with transverse field was used for theoretical
description of some other ferroelectric compounds, notably,
RbHSO$_4$~\cite{bib7, bib8} and NH$_4$HSO$_4$~\cite{bib8}
crystals.

In reference~\cite{bib9} the effect of hydrostatic pressure
on thermodynamic properties of the Rochelle salt was studied.
Hydrostatic pressure makes it possible to change the parameters of the
model.

It turns out that the Mitsui model covers not only the
ferroelectrics of the order-disorder type but also other physical
objects with two-minimum asymmetric potential. For instance, in
references~\cite{bib10} and~\cite{bib11} pseudospin-electron models
based on the Mitsui model were considered.

Despite a rather wide use of the Mitsui model, there is no
detailed analysis of its phase behavior. In reference~\cite{bib12}
the diagram of existence of the ferroelectric phase was
calculated. However, this diagram is far from being complete. More
detailed though still incomplete diagram was obtained in
reference~\cite{bib13}.

In the present paper, a rigorous and original mathematical
investigation of phase transitions in the Mitsui model is
proposed. We managed to construct a complete phase diagram of
the Mitsui model in the mean field approximation, first without
tunneling (or transverse field) and then with nonzero tunneling.
For all curves (surfaces) of the diagram the analytical
expressions, equations or systems of equations are given.

\section{Hamiltonian of the Mitsui model in the mean field approximation}
The Hamiltonian of the Mitsui model without external field reads
\begin{equation}
H=-\frac{1}{2}\sum\limits_{ij} J_{ij}
\left(S_i^{zA} S_j^{zA} +S_i^{zB} S_j^{zB}\right)
-\sum\limits_{ij} K_{ij} S_i^{zA} S_j^{zB}
-\Delta\sum\limits_i\left(S_i^{zA}-S_i^{zB}\right)
-\Omega\sum\limits_i\left(S_i^{xA}+S_i^{xB}\right).
\label{eq1}
\end{equation}
Here subscripts $A$ and $B$ denote two sublattices, $S_i^{\alpha
A}$ is $\alpha$-component of the pseudospin on $i$th site of
sublattice $A$; $J_{ij}$ and $K_{ij}$ are the interactions between
pseudospins of the same sublattice and of different
sublattices, respectively, $\Delta$ is the asymmetry of the
anharmonic potential; transverse field $\Omega$ describes the
tunneling between two wells of the two-minimum potential. The
model considered is a lattice one. However, since we use the mean
field approximation, we do not need to specify the type of the
lattice.

The transformations $S_i^{zB} \rightarrow -S_i^{zB}$, $K_{ij}
\rightarrow -K_{ij}$ transform Hamiltonian (\ref{eq1}) into the
Hamiltonian of the two-sublattice model with longitudinal field
$\Delta$ (the same for both sublattices) and transverse field
$\Omega$. Therefore, the two Hamiltonians are equivalent.

In the mean-field approximation, Hamiltonian (\ref{eq1}) reads
\begin{equation}
H=\frac{N}{2}\left\{
K\eta_{_A}\eta_{_B}+\frac{J}{2}\left(\eta_{_A}^2
+\eta_{_B}^2\right)\right\}+\sum\limits_i \left(H_i^A+H_i^B\right),
\label{eq2}
\end{equation}
where the following notations are introduced:
\begin{equation}
H_i^A=-\left(\Delta+K\eta_{_B}+J\eta_{_A}
\right)S_i^{zA}-\Omega S_i^{xA},
\label{eq3}
\end{equation}
\begin{equation}
H_i^B=-\left(-\Delta+K\eta_{_A}+J\eta_{_B}
\right)S_i^{zB}-\Omega S_i^{xB},
\label{eq4}
\end{equation}
\begin{equation}
K=\sum\limits_i K_{ij}=\sum\limits_j K_{ij}\,, \qquad
J=\sum\limits_i J_{ij}=\sum\limits_j J_{ij}\,, \label{eq5}
\end{equation}
$\eta_A=\langle S_i^{zA}\rangle$ and $\eta_B=\langle
S_i^{zB}\rangle$ are the average values of pseudospin on
sublattices $A$ and $B$, respectively, $N$ is the total number of
pseudospins.

\section{$\Omega =0$ case}

\subsection{Free energy, thermodynamic equilibrium conditions and order parameters}
Let us first consider  the \mbox{$\Omega=0$} case. In this case
the eigenvalues of Hamiltonians  $H_i^A$ and $H_i^B$ are as
follows:
\begin{equation}
\begin{array}{ll}
\displaystyle \lambda_{_A}=-\frac{1}{
2}\left(\Delta+K\eta_{_B}+J\eta_{_A} \right), & -\lambda_{_A}\,;
\\[2mm]
\displaystyle
\lambda_{_B}=-\frac{1}{2}\left(-\Delta+K\eta_{_A}+J\eta_{_B}
\right), & -\lambda_{_B}\,.
\end{array}
\label{eq6}
\end{equation}
The partition function for one unit cell reads
\begin{equation}
Z_i=\left(\re^{  -\beta\lambda_{_A}}+\re^{
\beta\lambda_{_A}}\right) \left(\re^{ -\beta\lambda_{_B}}+\re^{
\beta\lambda_{_B}}\right) \re^{
-\beta\left[K\eta_{_A}\eta_{_B}+\frac{J}{2}\left(\eta_{_A}^2
+\eta_{_B}^2 \right) \right]}\,. \label{eq7}
\end{equation}
The free energy per one unit cell is as follows:
\[
F=-\theta\ln{Z_i}\,,
\]
\begin{equation}
F=K\eta_{_A}\eta_{_B}+\frac{\displaystyle J}{\displaystyle
2}\left(\eta_{_A}^2+\eta_{_B}^2 \right) -\theta\ln{\left(\re^{
\displaystyle -\beta\lambda_{_A}}+\re^{ \displaystyle
\beta\lambda_{_A}}\right)}- \theta\ln{\left(\re^{ \displaystyle
-\beta\lambda_{_B}}+\re^{ \displaystyle
\beta\lambda_{_B}}\right)},
\end{equation}
where $\theta = {1}/{\beta}$ is thermodynamic temperature.
From thermodynamic equilibrium conditions
\begin{equation}
\begin{array}{l}
\displaystyle
\left(\frac{\partial{F}}{\partial{\eta_{_A}}}\right)_{T,\Delta}=0,
\\
\displaystyle
\left(\frac{\partial{F}}{\partial{\eta_{_B}}}\right)_{T,\Delta}=0
\end{array}
\label{eq9}
\end{equation}
we obtain the following equations for $\eta_{_A}$ and
$\eta_{_B}$\,:
\begin{equation}
\begin{array}{l}
2\eta_{_A}=\tanh{\left(-\beta\lambda_{_A}\right)},
\\
2\eta_{_B}=\tanh{\left(-\beta\lambda_{_B}\right)}.
\end{array}
\label{eq10}
\end{equation}
Taking into account equations (\ref{eq10}), we can rewrite the
expression for free energy in the following form:
\begin{equation}
F=K\eta_{_A}\eta_{_B}+ \frac{J}{2}\left(\eta_{_A}^2+ \eta_{_B}^2
\right) +\frac{\theta}{2}\ln
\left[{\left(\frac{1}{4}-\eta_{_A}^2\right)
\left(\frac{1}{4}-\eta_{_B}^2\right)}\right].
\label{eq11}
\end{equation}

The system of equations~(\ref{eq10}) has the solutions of two
kinds: for the first ones $\eta_{_A}=-\eta_{_B}$ (then the system
is reduced to one equation only), they exist for arbitrary
$\Delta$; for the second ones $\eta_{_A}\ne - \eta_{_B}$\,, they
exist in a bounded region of values of $\Delta$ and correspond to
the ferroelectric phase.

\begin{figure}[ht]
\begin{center}
\centerline{\includegraphics[width=0.38\textwidth]{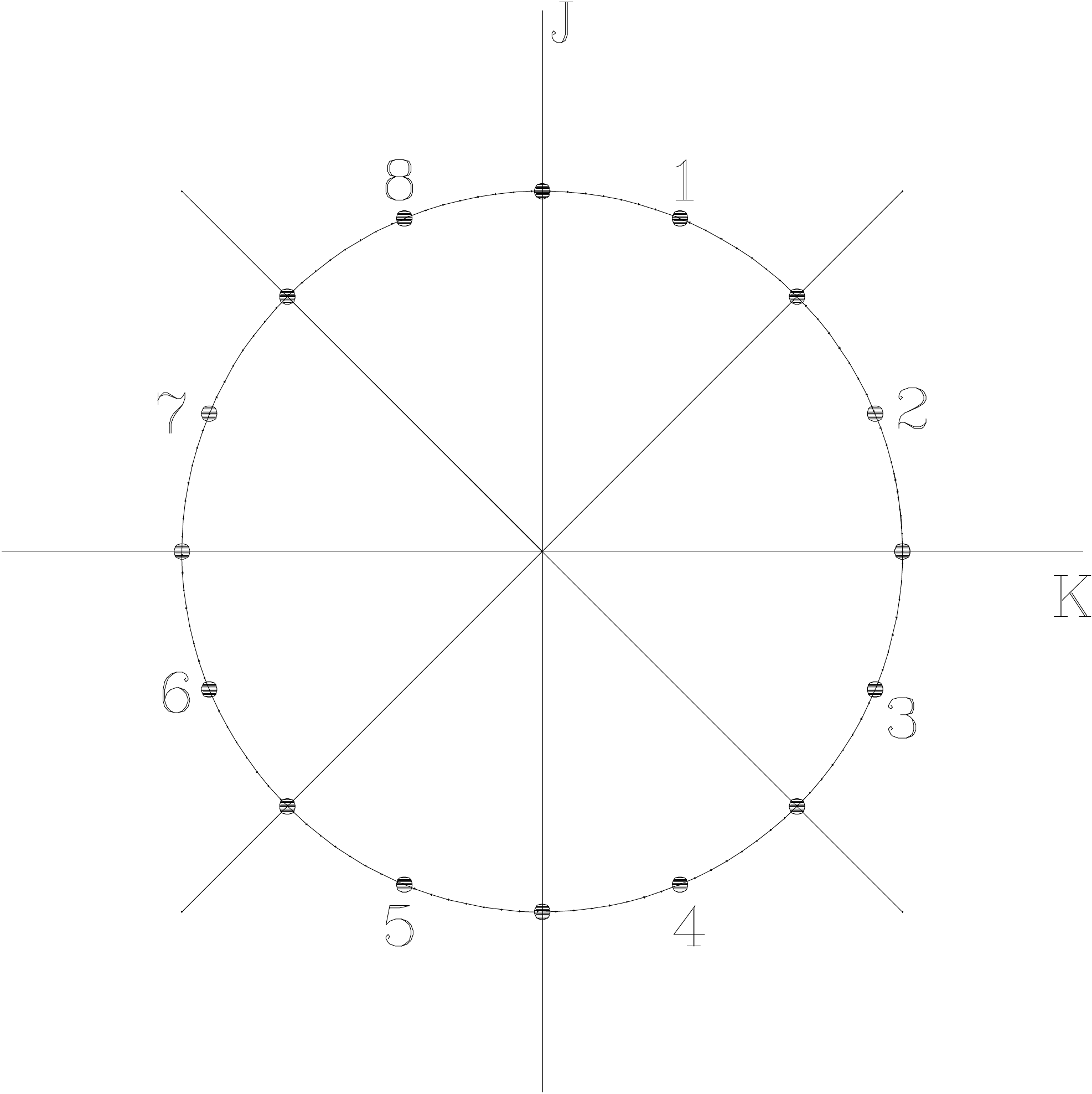}}
\caption{Division of the $(K,J)$ plane into eight segments, which
correspond to different relations between parameters $K$ and $J$.}
\label{fig1}
\end{center}
\end{figure}

\begin{figure}[!h]
\begin{center}
\centerline{\includegraphics[width=0.67\textwidth]{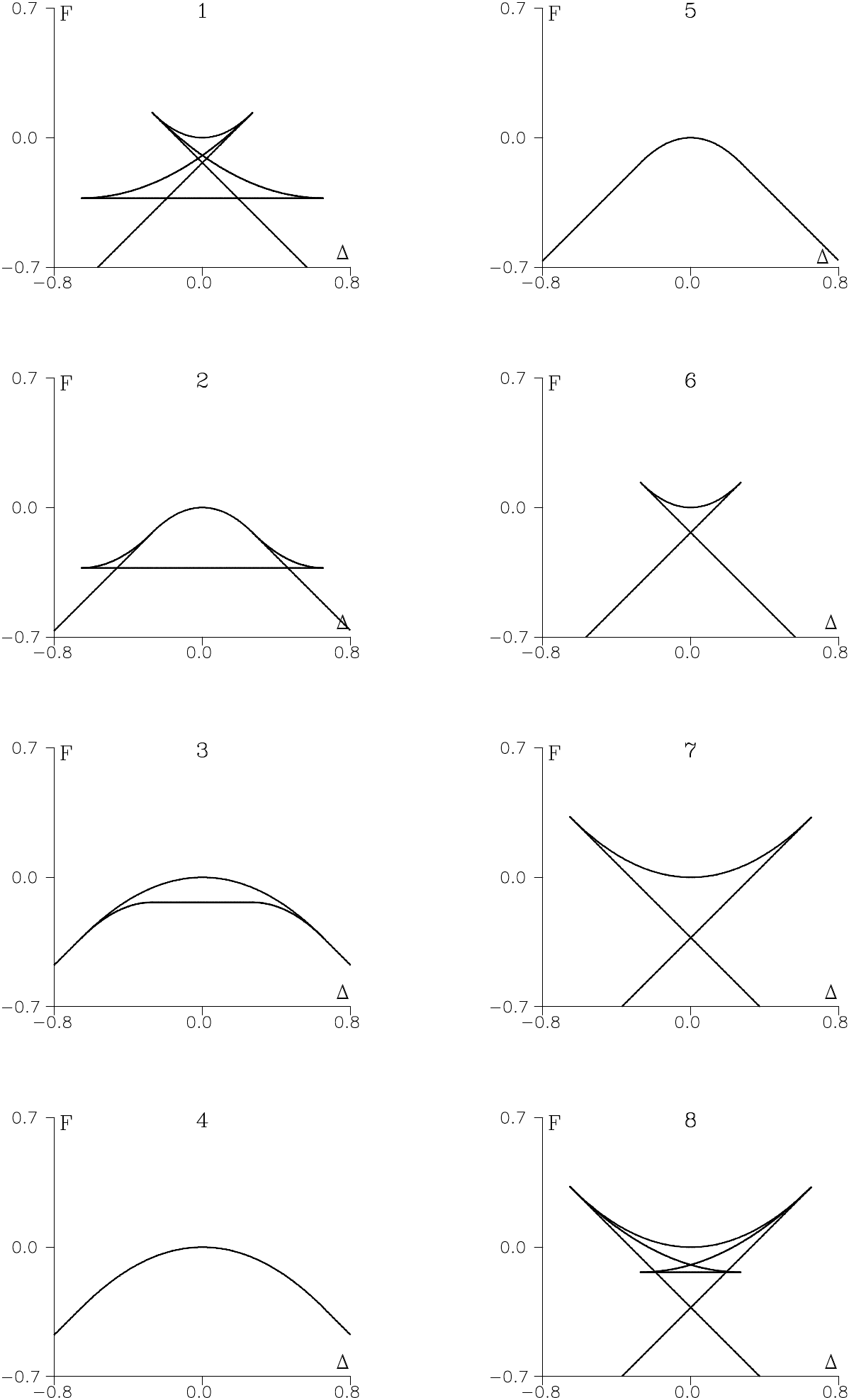}}
\caption[]{Free energy as a function of $\Delta$ at zero
temperature. Numbers over figures correspond to the unit-circle
points indicated in figure~\ref{fig1}.} \label{fig2}
\end{center}
\end{figure}

Let us divide the $(K,J)$ plane into eight segments as shown in
figure~\ref{fig1}. The plots of the free energy as a function of
$\Delta$ \emph{at zero temperature} in the centers of the
unit-circle arcs for every segment are shown in figure~\ref{fig2}.
(It is easy to obtain these plots from equations (\ref{eq10}) and
(\ref{eq11}) setting $\theta \rightarrow 0$ or $\beta \rightarrow
+\infty$.) As one can see from figure~2, in segments 4 and 5 there
are no phase transitions, in segments 6, 7, and 8 there are phase
transitions at $\Delta = 0$ only. In segment 3 there are only
second-order phase transitions that are easy to investigate. The
most complicated and interesting picture of phase transitions is
observed for segments 1 and 2, therefore we consider only the
region $K\geqslant 0$, $J\geqslant 0$.

Let us introduce new variables: ferroelectric order parameter $\xi
= \eta_{_A}+\eta_{_B}$ and antiferroelectric order parameter
$\sigma = \eta_{_A}-\eta_{_B}$\,. Let us also divide all energetic
values by $K+J$ and introduce the following notations:
\begin{equation}
a = \frac{K-J}{K+J}\,,\qquad  \gamma = \frac{\Delta}{K+J}\,,\qquad
t = \frac{\theta}{K+J}\,,\qquad f= \frac{F}{K+J}\,. \label{eq12}
\end{equation}
Now equations (\ref{eq10}) can be rewritten in the following form:
\begin{equation}
\begin{array}{rcl}
 \re^{
\xi/t}&=&\displaystyle\frac{(1+\xi)^2-\sigma^2}{(1-\xi)^2-\sigma^2}\,,
\\[4mm]
 \re^{  -a\sigma/t}&=&\re^{  -2\gamma/t}
\displaystyle\frac{(1+\sigma)^2-\xi^2}{(1-\sigma)^2-\xi^2}\,.
\end{array}
\label{eq13}
\end{equation}

If $\xi=0$ (first type solutions, both sublattices are equivalent
up to a sign of $\eta$) then the first equation becomes an
identity and the system (\ref{eq13}) is reduced to a single
equation. If $\xi\ne 0$ (second type solutions) then equations
(\ref{eq13}) can be rewritten in a simpler form:
\begin{equation}
\begin{array}{l}
\displaystyle \sigma=\pm\sqrt{1+\xi^2+2\xi\frac{1+\re^{ \xi/t}}
{1- \re^{ \xi/t}}}\,,
\\[4mm]
\displaystyle \gamma=\frac{a}{2}\sigma+\frac{\xi}{2}+
t\ln{\frac{1+\sigma-\xi}{1-\sigma+\xi}}\,.
\end{array}
\label{eq15}
\end{equation}
In view of the symmetry we consider only $\sigma\geqslant 0$,
$\xi\geqslant 0$ and $\gamma \geqslant 0$.

The expression for the free energy per unit cell in terms of new
variables reads
\begin{equation}
f = \frac{1}{4}\left(\xi^2-a\sigma^2\right)+
\frac{t}{2}\ln{\left\{\left[1-(\xi+\sigma)^2\right]
\left[1-(\xi-\sigma)^2\right]\right)}-2t\ln{2}.
\end{equation}

\subsection{Second-order phase transitions}

The second-order ferroelectric phase transition corresponds to the
branchpoint of the curve $\sigma(\gamma)$ ($a$ and $t$ being
fixed) or of the curve $\sigma (t)$ ($a$ and $\gamma$ being fixed)
where a solution of the first type turns into a solution of the
second type. We denote the value of $\sigma$ in this point by
$\tilde\sigma$. It is as follows:
\begin{equation}
\tilde\sigma = \lim\limits_{\xi\rightarrow 0}\sigma =
\sqrt{1-4t}\,. \label{eq17}
\end{equation}
One can see from this expression that the second-order phase
transitions exist up to the temperature
\begin{equation}
t_{\rm max}=\frac{1}{4}\,.
\end{equation}
Substituting $\xi = 0$ and $\sigma$ from equation (\ref{eq17}) in
equation (\ref{eq15}), we obtain the equation for the curve
$\gamma (t)$ of the second-order phase transitions:
\begin{equation}
\displaystyle \gamma=\frac{a}{2}\tilde\sigma+t \ln{
\frac{1+\tilde\sigma}{1-\tilde\sigma}}\,. \label{eq19}
\end{equation}

\begin{figure}[!b]
\includegraphics[width=0.45\textwidth]{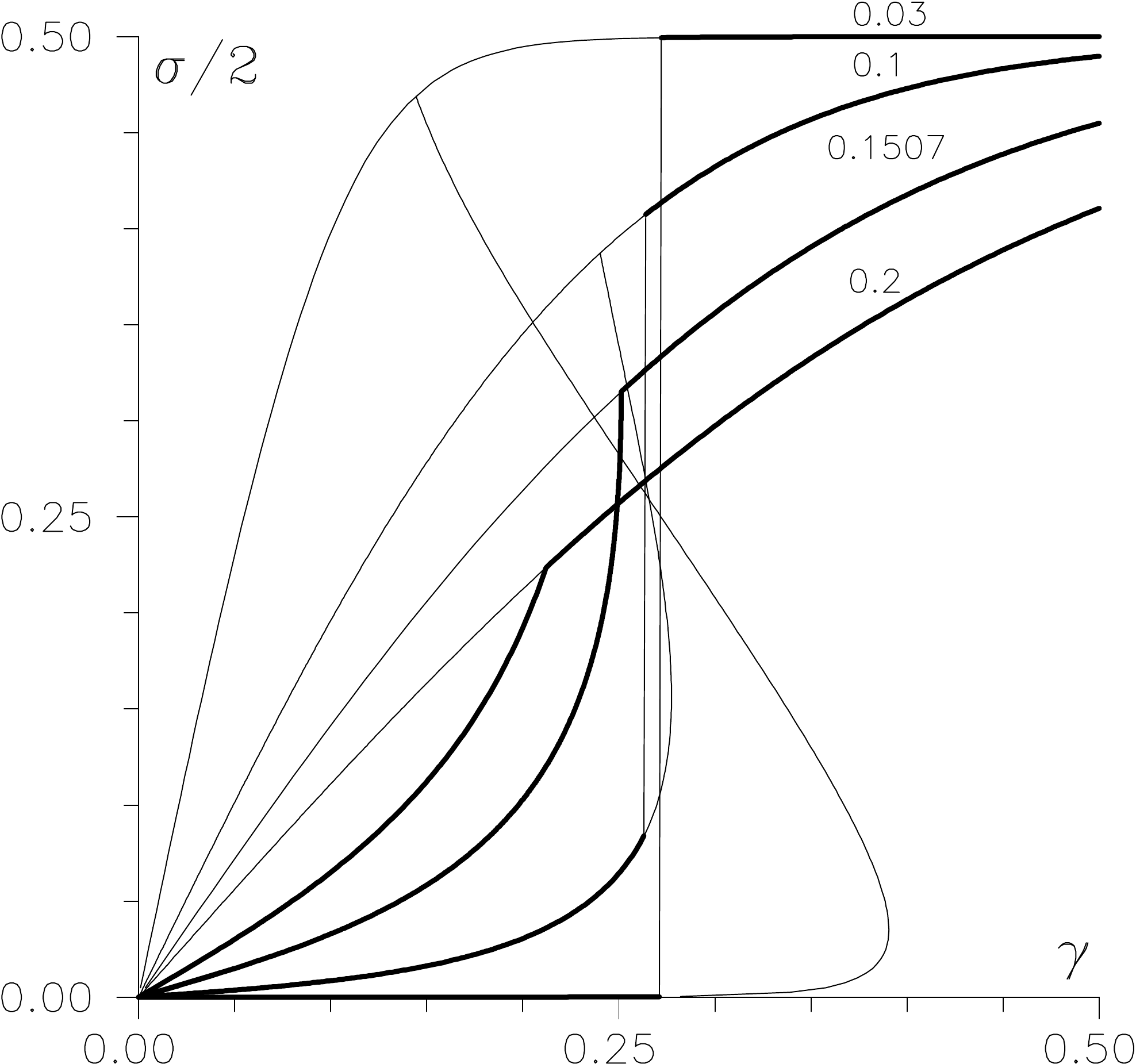}%
\hfill%
\includegraphics[width=0.45\textwidth]{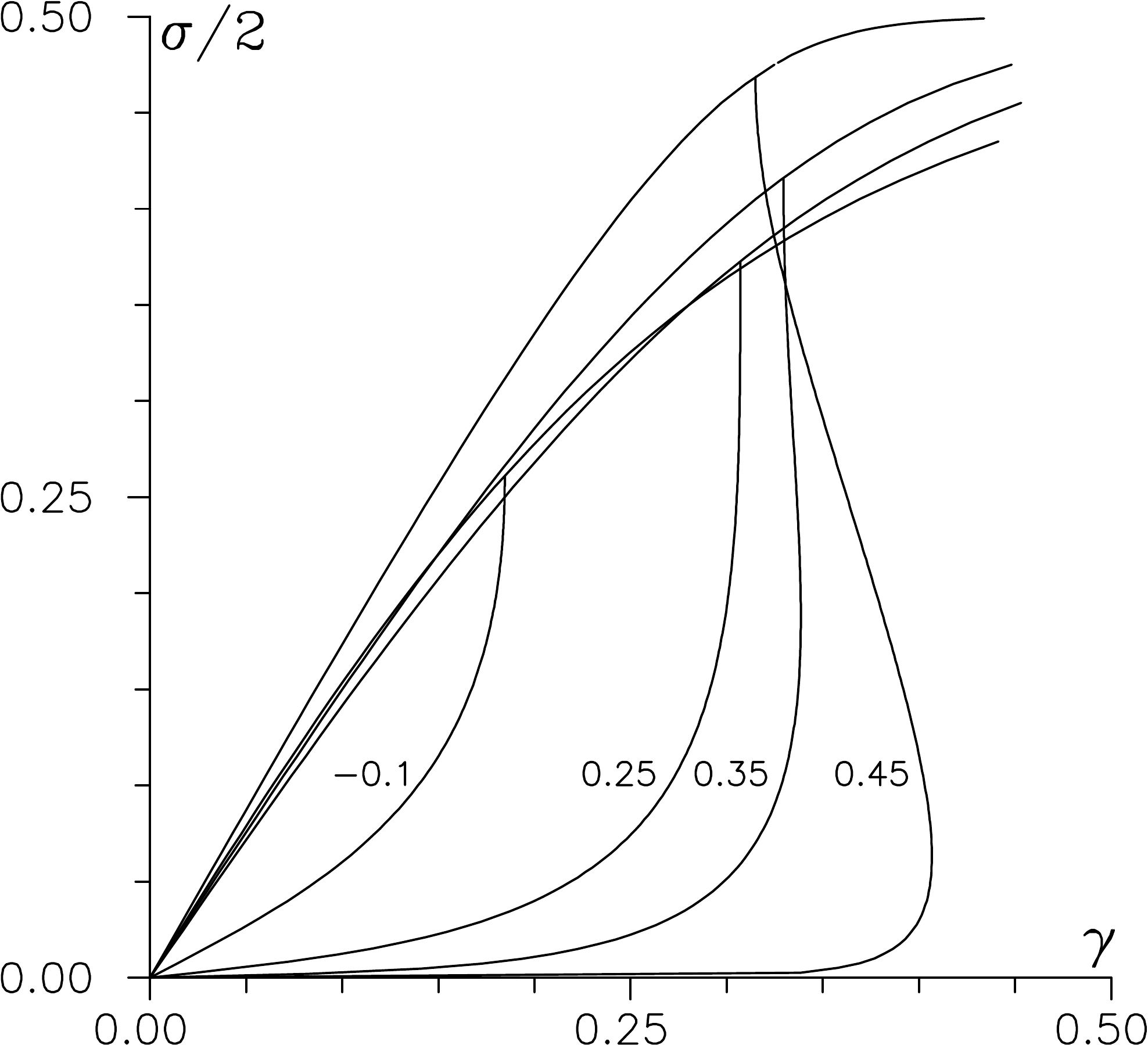}%
\\
\parbox[t]{0.49\textwidth}{%
\caption[]{Antiferroelectric order parameter as a function of
$\gamma$ for several values of temperature (numbers over the
curves). $a=0.0875$. Thermodynamically stable states are depicted
by heavy lines. At $t=0.03$ and $t=0.1$ there are first-order
phase transitions and at $t=0.1507$ and $t=0.2$ there are
second-order ones.} \label{fig3}
}%
\hfill%
\parbox[t]{0.49\textwidth}{%
\caption[]{Antiferroelectric order parameter as a function of
$\gamma$ for several values of parameter $a$ (numbers near the
curves). The values of the temperature are calculated using
expression (\ref{eq21}). At $a=-0.1$ and $a=0.25$ there are
second-order phase transitions and at $a=0.35$ and $a=0.45$ there
are first-order ones.} \label{fig4}}%
\end{figure}
%
Now let us find the minimal temperature for the existence of the
second-order phase transitions at fixed $a$. If $a = 1$, then only
second-order phase transitions exist. If $-1 < a < 1$, then there
are phase transitions of both second and first orders. The latter
exist from $t = 0$ to a certain value of the temperature. As one
can see from figure~\ref{fig3}, the tricritical point, i.e. the
point where the order of phase transition changes, can be
determine from the following condition:
\begin{equation}
\lim\limits_{\xi\rightarrow 0}\frac{\rd\gamma}{\rd\sigma}=0.
\end{equation}
We obtain for the tricritical temperature:
\begin{equation}
t_{\rm tc}=\frac{1}{3}+\frac{1}{6(a-1)}\,. \label{eq21}
\end{equation}

The tricritical point exists if, at the temperature determined by
equation (\ref{eq21}), the following condition is satisfied:
\begin{equation}
\lim\limits_{\xi\rightarrow 0}\frac{\rd^2\gamma}{\rd\sigma
^2}\leqslant 0 \qquad (\gamma\geqslant 0). \label{eq22}
\end{equation}
This is clear from figure~\ref{fig4} where curves ${ \sigma
(\gamma)}/{ 2}$ for several values of parameter $a$ and
corresponding values of the temperature [see equation
(\ref{eq21})] are depicted. The existence of the region where the
dependence $\sigma(\gamma)$ for $\xi\ne 0$ is two-valued indicates
that there is a first-order phase transition. This two-valuedness
disappears with decreasing $a$ and the order of the phase
transition changes. From equations (\ref{eq21}) and (\ref{eq22})
one obtains:
\begin{equation}
a\leqslant\frac{1}{4}\,.
\end{equation}

Let us find the maximum of curve $\gamma (t)$ at fixed $a$. The
extremum condition ${\rd\gamma}/{\rd t}=0$ and equation
(\ref{eq19}) yield:
\begin{equation}
\ln{\frac{1+\tilde\sigma}{1-\tilde\sigma}}=\frac{a+1}{\tilde\sigma}\,,
\qquad \gamma =
\frac{1}{4}\left(\frac{a+1}{\tilde\sigma}+(a-1)\tilde\sigma\right).
\label{eq24}
\end{equation}
Excluding $\tilde\sigma$, we obtain the equation for $\gamma$:
\begin{equation}
\left(2\gamma+\sqrt{(2\gamma)^2-a^2+1}\right)
\tanh\frac{2\gamma+\sqrt{(2\gamma)^2-a^2+1}}{2} = a+1.
\label{eq25}
\end{equation}

\subsection{First-order phase transitions}
Up to here we analyzed the second-order phase transitions and
found the expression for the surface $\gamma = \gamma (t,a)$ of
these transitions as well as the tricritical point and the
condition for its existence. Now let us consider the first-order
phase transitions.

A first-order phase transition corresponds to a point of
self-intersection of the curve for the free energy $f(\xi,\sigma)$
as a function of $\gamma$ at fixed $a$ and $t$ (or as a function
of $t$ at fixed $a$ and $\gamma$). But only this point of
self-intersection gives the first-order phase transition in which
the multivalued function for the free energy takes the minimal
value from all possible values at fixed $\gamma$.

If $-1\leqslant a\leqslant{1}/{4}$, then the first-order phase
transitions exists for $0\leqslant t < t_{\rm tc}$ [see equation
(\ref{eq21})]. The curve for them in the $(\gamma, t)$-plain (i.e,
the intersection points of the branches $\xi=0$ and $\xi\ne 0$ of
the free energy) can be found from the following system of
equations:
\begin{equation}
\begin{array}{l}
 \sigma=\sqrt{1+\xi^2+2\xi\displaystyle\frac{1+\re^{ \xi/t}} {1-  \re^{  \xi/t}}}\,,

\\[4mm]

\gamma=\displaystyle\frac{a}{2}\sigma+\frac{\xi}{2}+
t\ln{\frac{1+\sigma-\xi}{1-\sigma+\xi}}\,,

\\[4mm]
\gamma=\displaystyle\frac{a}{2}\sigma_1+t\ln{
\frac{1+\sigma_1}{1-\sigma_1}},

\\[4mm]
f(\xi,\sigma)=f(0,\sigma_1).
\end{array}
\label{eq27}
\end{equation}
If $t=0$, then the first-order phase transition occurs at
\begin{equation}
\gamma = \pm\frac{1}{4}(a+1). \label{eq27a}
\end{equation}
In the $t\ne0$ case the system of equations (\ref{eq27}) can be
solved only numerically.

\begin{figure}[ht]
\centerline{\includegraphics[width=0.65\textwidth]{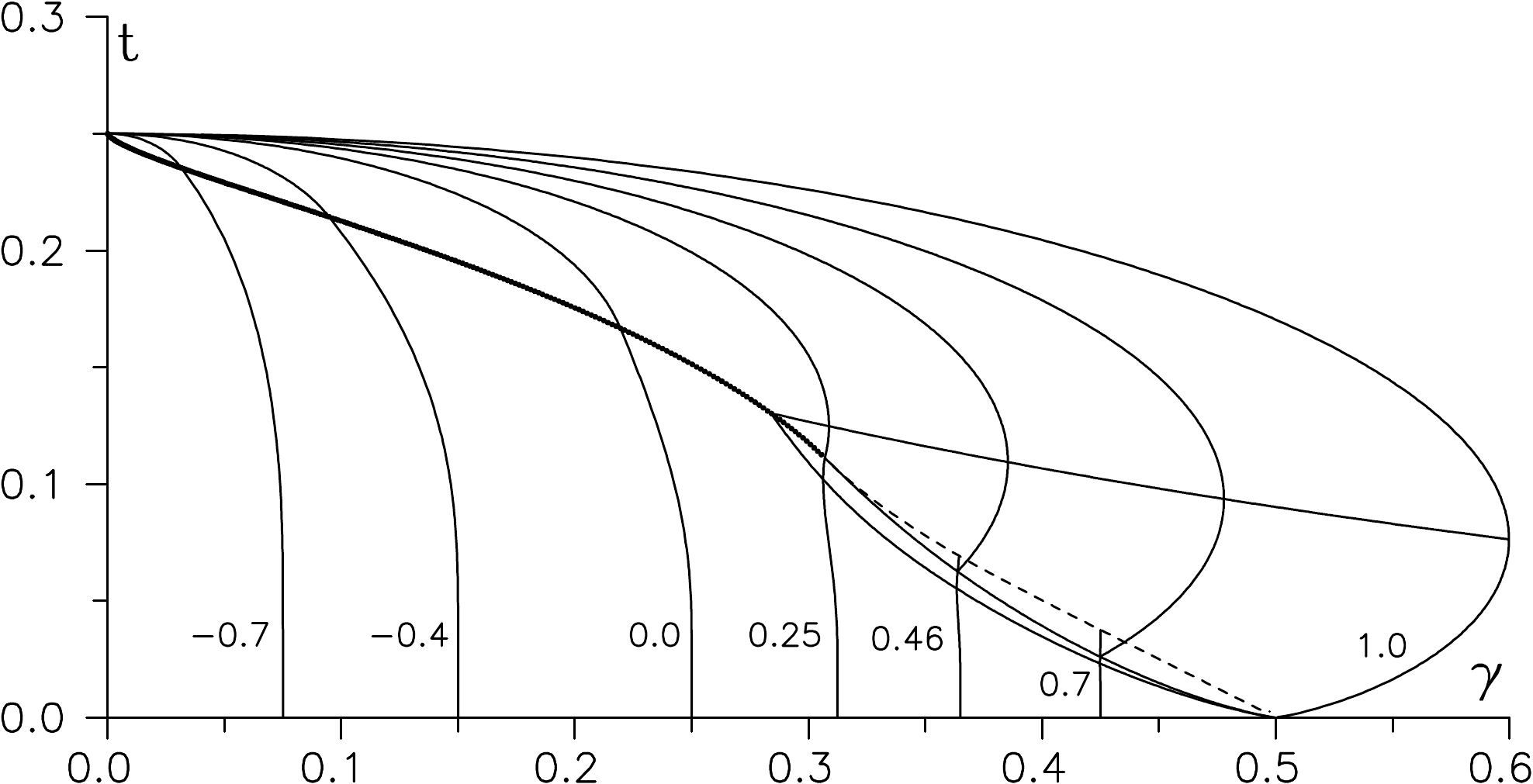}}
\caption[]{Coexistence curves in the $(\gamma,t)$-plane for
several values of parameter $a$ (numbers near the curves). The
curve of tricritical points (heavy line), the curve of minima for
the first-order phase transitions, the curve of maxima for the
second-order phase transitions, the curve of branchpoints and the
curve of critical points (dashed line) are indicated.}
\label{fig5}
\end{figure}

\begin{figure}
\centerline{\includegraphics[width=0.65\textwidth]{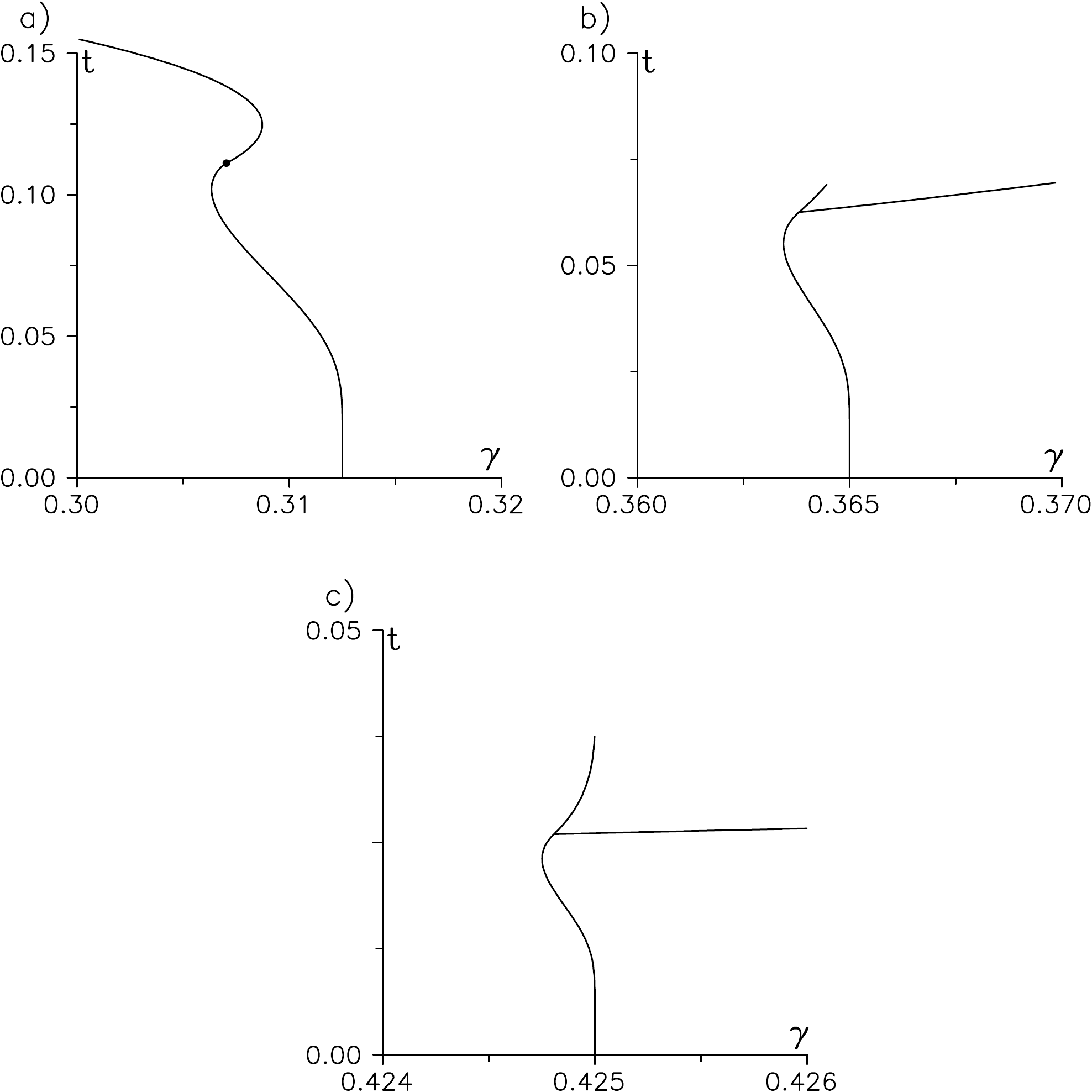}}
\caption[]{Fragments of the phase coexistence curves for a)
$a=0.25$ (the tricritical point is shown), b) $a=0.46$ and c)
$a=0.7$ (see figure~\ref{fig5}).} \label{fig6}
\end{figure}

The phase coexistence curves for several values of parameter $a$
are shown in figure~\ref{fig5}. The region of ferroelectric phase
is bounded by such a curve and by the coordinate axes. The curve
of tricritical points [more exactly, its projection on the plane
$(\gamma,t)$] is also shown in figure~\ref{fig5} (heavy line).
From this curve another two curves branch off: the curve of minima
of function $\gamma(t)$ for all possible values of $a$ and the
curve of maxima for the second-order phase transitions. The curve
of tricritical points furcates into the curve of branchpoints and
the curve of critical points (dashed line). If $a\gtrsim
0.1793293$ (four curves on the right), then there is a region with
two second-order phase transitions (see also figure~\ref{fig6}).
At this value of $a$ the maximum of the curve $\gamma(t)$ of the
second-order phase transitions coincides with the tricritical
point.

If $a\geqslant{1}/{4}$, then the upper part of the curve of the
first-order phase transitions corresponds to the phase transitions
within the ferroelectric phase. The temperature of these
transitions (at fixed $a$ and $\gamma$), i.e. the temperature for
the self-intersection points of the free energy curve, can be
found from the following system of equations (the solutions of the
type $\xi=\xi_1$ should be rejected):
\begin{equation}
\begin{array}{rcl}
 \sigma &=&\sqrt{1+\xi^2+2\xi\displaystyle\frac{1+\re^{
\xi/t}} {1-  \re^{  \xi/t}}}\,,
\\[4mm]
 \gamma &=&\displaystyle\frac{a}{2}\sigma+\frac{\xi}{2}+
t\ln{\frac{1+\sigma-\xi}{1-\sigma+\xi}}\,,
\\[3mm]
 \sigma_1 &=&\displaystyle \sqrt{1+\xi_1^2+2\xi_1\frac{1+\re^{
 \xi_1/t}} {1-  \re^{
\xi_1/t}}}\,,
\\[4mm]
 \gamma &=& \displaystyle\frac{a}{2}\sigma_1+\frac{\xi_1}{2}+
t\ln{\frac{1+\sigma_1-\xi_1}{1-\sigma_1+\xi_1}}\,,
\\[4mm]
&&f(\xi,\sigma)=f(\xi_1\,,\sigma_1).
\end{array}
\label{eq28}
\end{equation}

In figures~\ref{fig6}(b) and (c) some fragments of phase diagrams
in coordinates $(\gamma, t)$ for $a\geqslant{1}/{4}$ are shown.
The upper part of the curve of the first-order phase transitions
corresponds to the transitions within the ferroelectric phase.

If $a\geqslant{1}/{4}$, then the critical temperature can be
determined from the following system of equations:
\begin{equation}
\begin{array}{l}
\displaystyle \frac{\rd\gamma}{\rd\sigma}=0,
\\[3mm]
\displaystyle \frac{\rd^2\gamma}{\rd\sigma ^2}=0,
\\[3mm]
 \sigma=\displaystyle\sqrt{1+\xi^2+2\xi\frac{1+\re^{
\xi/t}} {1-  \re^{  \xi/t}}}\,,
\\[4mm]

\gamma=\displaystyle\frac{a}{2}\sigma+\frac{\xi}{2}+
t\ln{\frac{1+\sigma-\xi}{1-\sigma+\xi}}\,.
\end{array}
\label{eq29}
\end{equation}
This system of equations can be rewritten in the form:
\begin{equation}
\begin{array}{rcl}
\sigma^2&=&1+\xi^2-2t+\displaystyle\frac{ 2}{
a}\left(t-\sqrt{a^2\xi^2+2at\xi^2(1-a)+t^2(1+a)^2}\right),

\\[3mm]
\sigma^2&=&1+a\xi^2-3t + \displaystyle\frac{  1}{
a}\left(t-\sqrt{\left[a\xi^2 (1-a)+t(1+a)\right]^2
-4a^2\xi^2(2at-2t-a)}\right),

\\[3mm]
 \sigma^2&=&\displaystyle1+\xi^2+2\xi\frac{1+\re^{
\xi/t}} {1- \displaystyle \re^{  \xi/t}}\,,

\\[4mm]
 \gamma&=&\displaystyle\frac{a}{2}\sigma+\frac{\xi}{2}+
t\ln{\frac{1+\sigma-\xi}{1-\sigma+\xi}}\,,
\end{array}
\label{eq30}
\end{equation}
whence it follows
\begin{equation}
 \xi^2=\frac{-q-[(3a-1)t-a]
\sqrt{q-2(a+1)t\left[(a-1)^2t-a^2\right]}}{2a(a-1)[2(a-1)t-a]}\,,
\label{eq31}
\end{equation}
where $q=(a-1)(3a^2-14a-1)t^2-2a(2a^2-5a+1)t+a^2(a-1)$, and hence
the system of equations is reduced to a single transcendental
equation.

As one can see from figure~\ref{fig6}, the curve $\gamma (t)$ of
the first-order phase transitions has a minimum if $a$ is big
enough. To find it, let us differentiate the last equation of
system (\ref{eq27}) with respect to $t$ and then substitute
${\rd\gamma}/{\rd t}=0$. After simple transformations we obtain
the following equation:
\begin{equation}
\xi^2-a\sigma^2+4\gamma\sigma=-a\sigma_1^2+4\gamma\sigma_1\,,
\label{eq32}
\end{equation}
which together with equations (\ref{eq27}) gives the point of
minimum. At zero temperature the solutions of equations
(\ref{eq27}) also satisfy it:
\begin{equation}
\lim\limits_{t\rightarrow 0}\frac{ \rd\gamma}{ \rd t}=0.
\end{equation}
The branchpoint of the curve (at fixed
$a\geqslant\displaystyle{1}/{4}$) can be found from equations
(\ref{eq27}), substituting
\begin{equation}
t=\displaystyle\frac{1-\sigma_1^2}{4}\,. \label{eq34}
\end{equation}

\subsection{Regions of existence of the ferroelectric phase}
\begin{figure}[!b]
\includegraphics[width=0.45\textwidth]{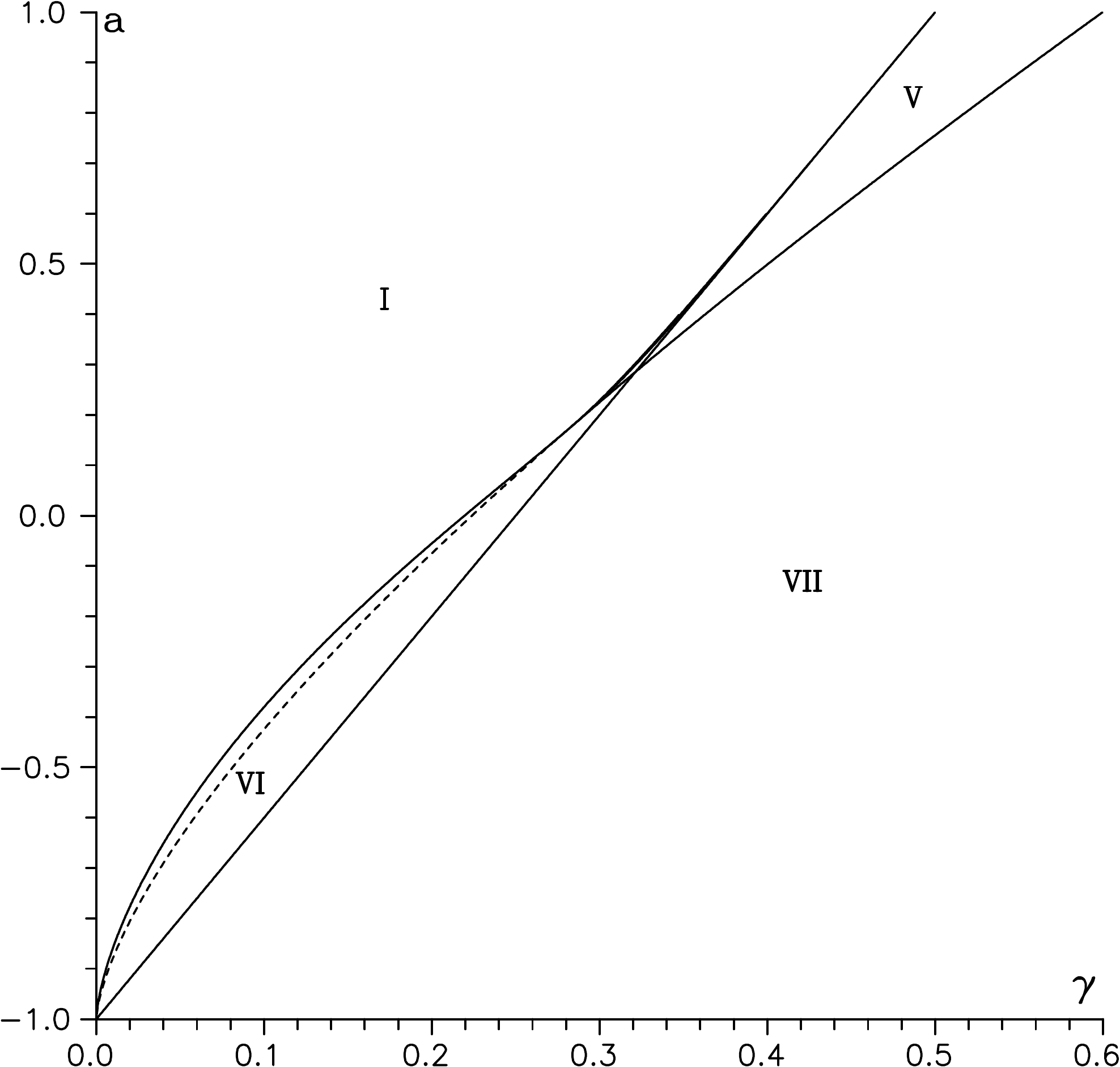}%
\hfill%
\includegraphics[width=0.46\textwidth]{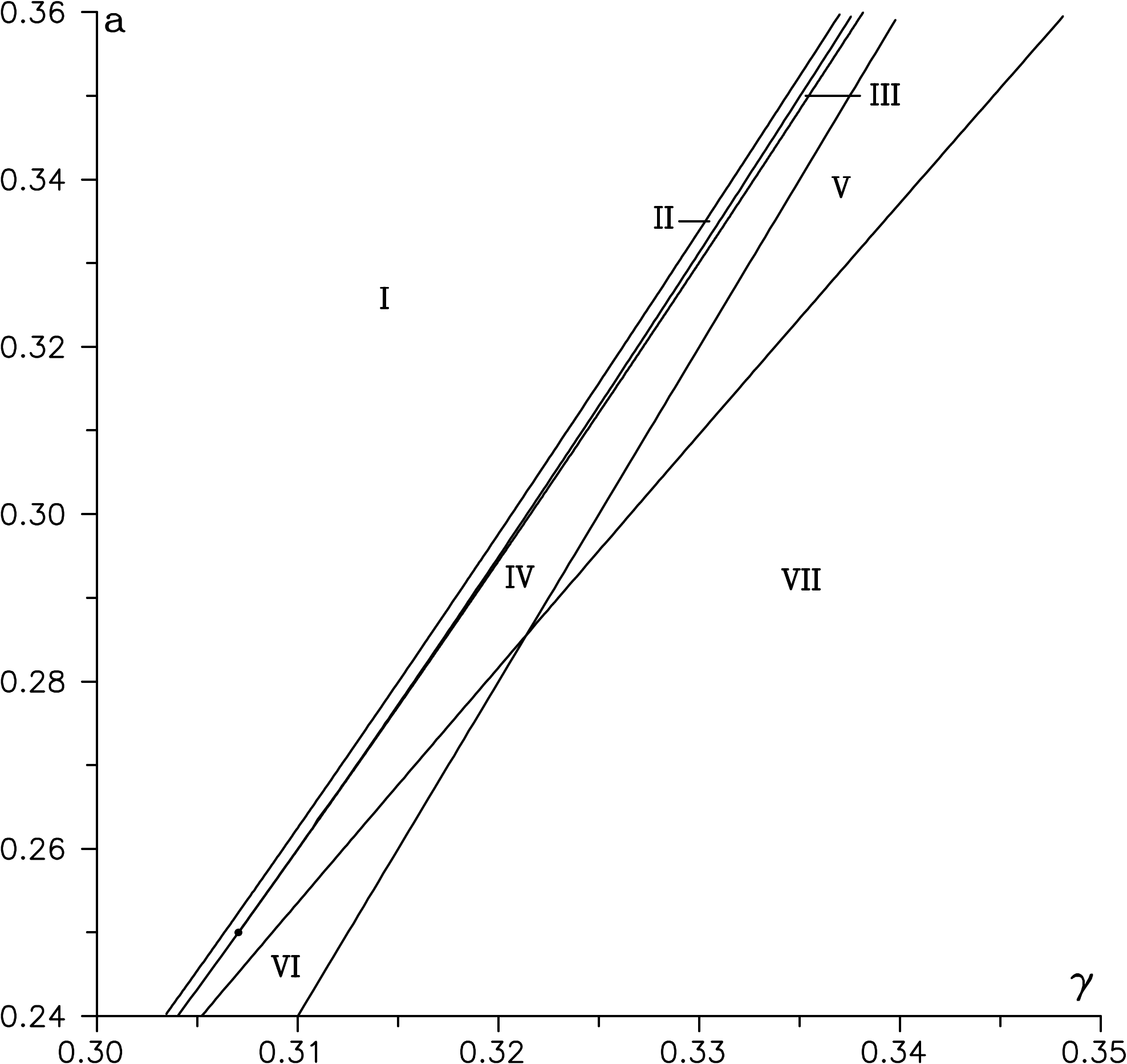}%
\\
\parbox[t]{0.49\textwidth}{%
\caption[]{Regions of existence of the ferroelectric phase (see
also figures~\ref{fig9} and~\ref{fig10}). Dashed line
(prolongation of the line which bounds region V) is shown only for
comparison with the diagram from~\cite{bib12}.} \label{fig7}
}%
\hfill%
\parbox[t]{0.49\textwidth}{%
\caption[]{Regions of existence of the ferroelectric phase
(fragment). The filled circle is the branchpoint.} \label{fig8}
}%
\end{figure}
To conclude, we obtained explicit expressions or system of
equations for all special points of the phase curve at fixed $a$.
This makes it possible to construct a diagram of the regions
where the ferroelectric phase exists. There are seven regions in
the $(\gamma, a)$ plane (figures~\ref{fig7} and~\ref{fig8}). They
are
bounded by the following curves:\\
(1) the curve of minima for the first-order phase transitions [equation (\ref{eq32})];\\
(2) the curve of tricritical points [equations (\ref{eq17}), (\ref{eq19}), and (\ref{eq21})];\\
(3) the curve of branchpoints [equations (\ref{eq27}), and (\ref{eq34})];\\
(4) the curve of critical points [equation (\ref{eq30})];\\
(5) the curve of maxima for the second-order phase transitions
[equation (\ref{eq25})];\\
(6) the straight line of the first-order phase transitions at zero
temperature [equation (\ref{eq27a})].


In the point ($a\approx 0.179329$, $\gamma=0.283995$) the curve of
minima for the first-order phase transitions and the curve of
maxima for the second-order phase transitions branch off from the
curve of tricritical points and in the point ($a=0.25$,
$\gamma=0.307041$) the curve of critical points branches out into
the curve of branchpoints and the curve of critical points. All
curves except for the curve of maxima for the second-order phase
transitions converge at $a=1$.
\begin{figure}[h!]
\centerline{\includegraphics[width=0.6\textwidth]{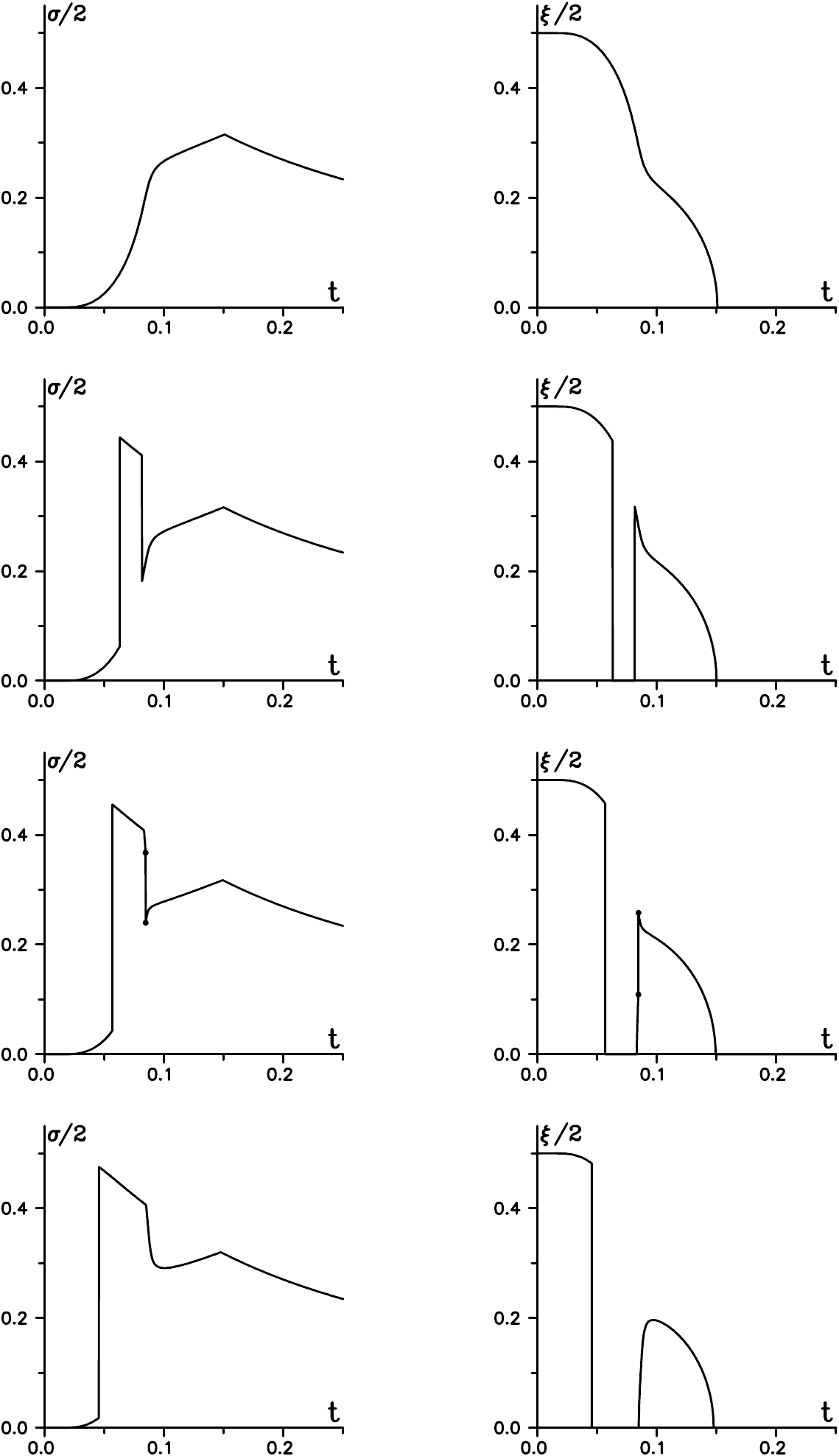}}
\caption[]{Parameters of ferroelectric and antiferroelectric
ordering for different regions where the ferroelectric phase
exists. Only thermodynamically stable states are depicted.
$a=0.36$. From top to bottom: I) $\gamma=0.337$; II)
$\gamma=0.3375$; III) $\gamma=0.338$; and IV)~$\gamma=0.339$. For
region III the phase transition within the ferroelectric phase is
indicated by filled circles.} \label{fig9}
\end{figure}
\begin{figure}[!h]
\centerline{\includegraphics[width=0.65\textwidth]{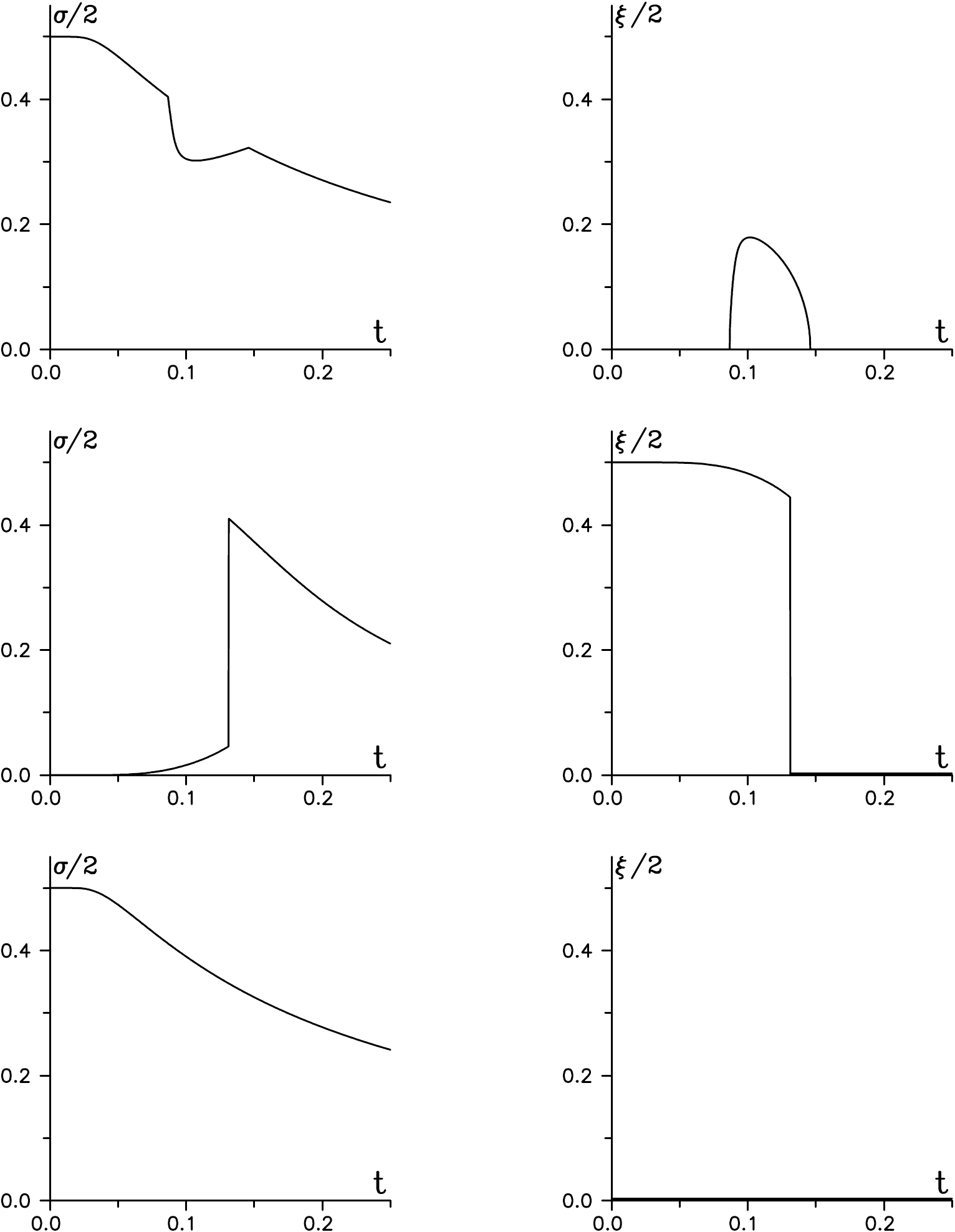}}
\caption[]{ (Continuation of figure~\ref{fig9}.) V) $a=0.36$,
$\gamma=0.34$; VI) $a=-0.4$, $\gamma=0.14$; VII)~$a=0.36$,
$\gamma=0.35$.} \label{fig10}
\end{figure}

In figures~\ref{fig9} and~\ref{fig10}, the behavior of parameters
$\sigma$ and $\xi$ is shown for every region. In region I (figures
\ref{fig7} and~\ref{fig8}) only one high-temperature second-order
phase transition exists. In region II, in addition to the
mentioned one, there are two first-order phase transitions. In
narrow region III there are four phase transition: two of the
first and two of the second order; one first-order phase
transition is within the ferroelectric phase. In region IV there
are three phase transitions, one of them being the first-order
transition. In region V there are two second-order phase transitions just like
for the Rochelle salt. In region VI there is only one phase
transition (of the first order) and in region VII there are no
phase transitions at all.

\section{$\Omega \ne 0$ case}
\subsection{Free energy and conditions for thermodynamic equilibrium}

Now let us consider the case of nonzero tunneling. The eigenvalues
of one-site Hamiltonians (\ref{eq3}) and (\ref{eq4}) are as
follows:
\begin{equation}
\begin{array}{l}
\lambda_{_A}=-\displaystyle\frac{ 1}{ 2}\sqrt{\Delta_{_A}^2+
\Omega^2}\,, \qquad-\lambda_{_A}\,,\qquad
\Delta_{_A}=\Delta+K\eta_{_B}+J\eta_{_A}\,;
\\[3mm]
\lambda_{_B}=-\displaystyle\frac{ 1}{ 2}
\sqrt{\Delta_{_B}^2+\Omega^2}\,, \qquad-\lambda_{_B}\,,\qquad
\Delta_{_B}=-\Delta+K\eta_{_A}+J\eta_{_B}\,;
\end{array}
\label{eq}
\end{equation}
and the free energy per one unit cell reads
\begin{eqnarray}
F&=&K\eta_{_A}\eta_{_B}+\frac{ J}{ 2}\left(\eta_{_A}^2+\eta_{_B}^2
\right) -\frac{1}{2}\left(\sqrt{\Delta_{_A}^2+\Omega^2
}+\sqrt{\Delta_{_B}^2+\Omega^2 }\right)-\nonumber
\\
&-&\theta\ln{\left(1+\re^{ -\beta\sqrt{\Delta_{_A}^2+\Omega^2
}}\right)}- \theta\ln{\left(1+\re^{
-\beta\sqrt{\Delta_{_B}^2+\Omega^2 }}\right)}. \label{eqa}
\end{eqnarray}
The conditions of thermodynamic equilibrium (\ref{eq9}) yield
the following equations:
\begin{equation}
\begin{array}{l}
2\eta_{_A}= \displaystyle
\frac{\Delta_{_A}}{\sqrt{\Delta_{_A}^2+\Omega^2}}
\tanh{\left(-\beta\lambda_{_A}\right)},
\\[4mm]
2\eta_{_B}= \displaystyle
\frac{\Delta_{_B}}{\sqrt{\Delta_{_B}^2+\Omega^2}}
\tanh{\left(-\beta\lambda_{_B}\right)}.
\end{array}
\label{eqo3}
\end{equation}
Like in the $\Omega=0$ case, let us pass to dimensionless
values, introducing one more notation:
\begin{equation}
\omega=\frac{\Omega}{K+J}\,. \label{eqb}
\end{equation}
Let us also introduce new variables:
\begin{equation}
\begin{array}{l}
x= \displaystyle
\frac{\Delta_{_A}+\Delta_{_B}}{2(K+J)}=\frac{\eta_{_A}+\eta_{_B}}{2}=
\frac{\xi}{2}\,,
\\[3mm]
y= \displaystyle
\frac{\Delta_{_A}-\Delta_{_B}}{2(K+J)}=\gamma-a\frac{\eta_{_A}-
\eta_{_B}}{2}=\gamma-a\frac{\sigma}{2}\,.
\end{array}
\label{eqc}
\end{equation}
Then the system of equations (\ref{eqo3}) becomes:
\begin{equation}
\begin{array}{l}
2x=A(x+y)-A(-x+y),
\\[2mm]
\displaystyle
\frac{2(\gamma-y)}{a}=A(x+y)+A(-x+y),
\end{array}
\label{eqo6}
\end{equation}
where
\begin{equation}
A(z)=\frac{z}{2\sqrt{z^2+\omega^2}}
\tanh{\left(\frac{\sqrt{z^2+\omega^2}}{2t}\right)},
\label{eq07}
\end{equation}
and the expression for free energy takes the form:
\begin{eqnarray}
f&=&x^2- \frac{(\gamma-y)^2}{a}-\frac{1}{2}
\left(\sqrt{(x+y)^2+\omega^2}+\sqrt{(x-y)^2+\omega^2}\right)-\nonumber
\\[3mm]
&-&t\ln{\left(1+\re^{ -{\sqrt{(x+y)^2+\omega^2}}/{t}}\right)}
-t\ln{\left(1+\re^{ -{\sqrt{(x-y)^2+\omega^2}}/{t}}\right)}.
\end{eqnarray}

\subsection{Second-order phase transitions}
Like in the $\Omega=0$ case, the system of equations (\ref{eqo6}) has
solutions of two types: 1) $x=0$; then the first equation becomes
an identity; and 2) $x\ne 0$, which corresponds to the
ferroelectric phase. Letting $x$ tend to zero, we obtain from
(\ref{eqo6}) the system of equations for hypersurface
$\gamma=\gamma(\omega,a,t)$ of the second-order phase transitions:
\begin{equation}
\begin{array}{l}
B(\tilde y)-1=0,
\\[2mm]
\gamma=\tilde y+aA(\tilde y),
\end{array}
\label{eqo8a}
\end{equation}
where $B(z)={\rd A(z)}/{\rd z}$, $\tilde y= \lim
\limits_{x\rightarrow 0}y$.  Having  determined $B(z)$, we can
write the system in the form:
\begin{equation}
\begin{array}{l}
\displaystyle s^2- \frac{2t\omega^2s}{{\tilde y}^2\sqrt{\tilde
y^2+\omega^2}}+  \frac{4t\left({\tilde
y}^2+\omega^2\right)}{\tilde y^2}-1=0,
\\[4mm]
\displaystyle \gamma=\tilde y+\frac{as\tilde y}{2\sqrt{\tilde
y^2+\omega^2}}\,,
\end{array}
\label{eqo8}
\end{equation}
where the following notation is introduced:
\[
s=\tanh{\left(\frac{\sqrt{\tilde y^2+\omega^2}}{2t}\right)}.\]
Setting $\gamma$ equal to zero in equation (\ref{eqo8}), we obtain
the expression for maximal temperature at which the phase
transitions exist:
\begin{equation}
t_{\rm max}=  \frac{\omega}{\ln{ \displaystyle
\frac{1+2\omega}{1-2\omega}}}\,. \label{eqo9}
\end{equation}
The logarithm in the latter expression makes sense if
$\omega\leqslant {1}/{2}$. Hence,
\begin{equation}
\omega_{\rm max}= \frac{1}{2}\,. \label{eqo10}
\end{equation}
Differentiating equations (\ref{eqo8}) with respect to $t$,
setting ${\rd\gamma}/{\rd t}$ equal to zero, and eliminating
${\rd\tilde y}/{\rd t}$, we obtain after simple transformations:
\begin{align}
&2(1+a)\left(\tilde y^2+\omega^2\right) \left(\tilde
y^2s\sqrt{\tilde y^2+\omega^2}-t\left(\tilde
y^2+\omega^2\right)\right)\nonumber
\\&
=a\left(3t\omega^2s\sqrt{\tilde y^2+\omega^2}- 2\tilde
y^2s\sqrt{\tilde y^2+\omega^2}+ 6t\omega^2\left(\tilde
y^2+\omega^2\right)-\omega^2\tilde y^2s^2\right). \label{eqo10a}
\end{align}
This equation together with equations (\ref{eqo8}) determines the
point of maximum for the curve $\gamma(t)$ of the second-order
phase transitions (at fixed $\omega$ and $a$).

The equation for tricritical point is similar to that in the
$\Omega=0$ case:
\begin{equation}
\lim\limits_{x\rightarrow 0} \frac{\rd\gamma}{\rd y}=0.
\label{eqo11}
\end{equation}
From this equation we obtain:
\begin{equation}
(a+1)D\left(\tilde y\right)-3a\left[C\left(\tilde
y\right)\right]^2=0, \label{eqo12}
\end{equation}
where the following notations are introduced:
\begin{equation}
C(z)=\frac{\rd^2A(z)}{\rd z^2}\,, \qquad D(z)=\frac{\rd^3A(z)}{\rd
z^3}\,. \label{eqo13}
\end{equation}
The tricritical point exists if both equation (\ref{eqo11}) and
the following condition are satisfied:
\begin{equation}
\lim\limits_{x\rightarrow 0} \frac{\rd^2\gamma}{\rd y^2}\leqslant
0\qquad(\gamma\geqslant 0). \label{eqo14}
\end{equation}
This yields the equation
\begin{equation}
8\left[D\left(\tilde y\right)\right]^3 -9C\left(\tilde
y\right)D\left(\tilde y\right)E\left(\tilde y\right)+
\frac{9}{5}\left[C\left(\tilde y\right)\right]^2 F\left(\tilde
y\right)=0, \label{eqo15}
\end{equation}
where
\begin{equation}
E(z)=\frac{\rd^4A(z)}{\rd z^4}\,, \qquad F(z)=\frac{\rd^5A(z)}{\rd
z^5}\,. \label{eqo16}
\end{equation}

If the second-order phase transitions exist until $t=0$, then it
follows from equations (\ref{eqo8}), that for this
zero-temperature transition $\gamma$ is the following:
\begin{equation}
\gamma=(2\omega)^{-{2}/{3}}\left(a+(2\omega)^{{2}/{3}}
\right)\tilde y, \label{eqo17}
\end{equation}
where
\begin{equation}
\tilde y= \frac{1}{2}(2\omega)^{{2}/{3}}
\left(1-(2\omega)^{{2}/{3}}\right)^{{1}/{2}}. \label{eqo17a}
\end{equation}

Rewriting equation (\ref{eqo12}) for zero temperature we obtain:
\begin{equation}
a=\frac{5(2\omega)^{{2}/{3}}-4}{4(2\omega)^{{2}/{3}}-5}\,.
\label{eqo18}
\end{equation}
If $a$ is bigger than this value (at fixed $\omega$), then the
second-order phase transitions begin at zero temperature. It
follows from equation (\ref{eqo15}) that equation (\ref{eqo18}) is
satisfied under condition
\begin{equation}
\omega \geqslant 2^{-{5}/{2}}\approx 0.1768;
\end{equation}
then $a\leqslant{1}/{2}$.

\subsection{First-order phase transitions}
The points of the first-order ferroelectric phase transitions can
be calculated from the following system of equations:
\begin{equation}
\begin{array}{rcl}
2x&=&A(x+y)-A(-x+y),
\\[2mm]
\displaystyle \frac{2(\gamma-y)}{a}&=&A(x+y)+A(-x+y),
\\[2mm]
\displaystyle \frac{\gamma-y_1}{a}&=&A(y_1),
\\[2mm]
f(x,y)&=&f(0,y_1).
\end{array}
\label{eqo19}
\end{equation}
Differentiating the last equation with respect to $t$ and
eliminating the derivatives, we obtain the following equation:
\begin{eqnarray}
x^2&-&\frac{(\gamma-y)^2}{a}-\left(x+\frac{\gamma-y}{a}\right)
\frac{(x+y)^2+\omega^2}{x+y}\nonumber
\\
&-&\left(x-\frac{\gamma-y}{a}\right)
\frac{(x-y)^2+\omega^2}{x-y}=-\frac{(\gamma-y_1)^2}{a}
-\frac{2(\gamma-y_1)}{a}\frac{y_1^2+\omega^2}{y_1}\,,
\label{eqo20}
\end{eqnarray}
which, together with the system of equations~(\ref{eqo19}), yields
the extremum of the curve $\gamma(t)$ of the first-order phase
transitions (at fixed $\omega$). If $t=0$, then the solutions of
the system of equations~(\ref{eqo19}) satisfy
equation~(\ref{eqo20}) as well. Hence, for the curve
$\gamma(t)$ of the first-order phase transitions we have
\begin{equation}
\lim\limits_{t\rightarrow 0}\frac{\rd\gamma}{\rd t}=0.
\end{equation}

Like in the $\Omega=0$ case, we can obtain the system of equations
for the tricritical point setting ${\rd\gamma}/{\rd y}$ and
${\rd^2\gamma}/{\rd y^2}$ equal to zero:
\begin{equation}
\begin{array}{l}
 a=\displaystyle\frac{2-B_{-}-B_{+}}{2B_{-}B_{+}-B_{-}-B_{+}}\,,
\\[3mm]
C_{+}\left(1-B_{-}\right)^3+C_{-}\left(1-B_{+}\right)^3=0,
\\[2mm]
2x=A_{+}-A_{-}\,,
\\[2mm]
\displaystyle \frac{2(\gamma-y)}{a}=A_{+}+A_{-}\,,
\end{array}
\label{eqo21}
\end{equation}
where $A_\pm=A(\pm x+y)$, $B_\pm=B(\pm x+y)$, and $C_\pm=C(\pm
x+y)$.

Rewriting the system of equations (\ref{eqo21}) for zero temperature,
we obtain:
\begin{equation}
\begin{array}{l}
\displaystyle
\frac{z_{+}\left(2\left(z_{-}^2+\omega^2\right)^{{3}/{2}}-
\omega^2\right)^3}{\left(z_{-}^2+\omega^2\right)^2}+
\frac{z_{-}\left(2\left(z_{+}^2+\omega^2\right)^{{3}/{2}}-
\omega^2\right)^3}{\left(z_{+}^2+\omega^2\right)^2}=0,
\\[5mm]
\displaystyle
\frac{z_{+}}{2\sqrt{z_{+}^2+\omega^2}}
-\frac{z_{-}}{2\sqrt{z_{-}^2+\omega^2}}-z_{+}+z_{-}=0,
\\[5mm]
\displaystyle
a=\frac{4\left(z_{+}^2+\omega^2\right)^\frac{3}{2}
\left(z_{-}^2+\omega^2\right)^\frac{3}{2}-\omega^4}
{\omega^2\left( \omega^2-\left(z_{+}^2+\omega^2\right)^\frac{3}{2}
-\left(z_{-}^2+\omega^2\right)^\frac{3}{2} \right)}+1,
\\[5mm]
\displaystyle
\gamma=\frac{a}{4}\left(\frac{z_{+}}{\sqrt{z_{+}^2+\omega^2}}+
\frac{z_{-}}{\sqrt{z_{-}^2+\omega^2}}\right)
+\frac{1}{2}\left( z_{+}+z_{-}\right),
\end{array}
\label{eqo22}
\end{equation}
where $z_\pm=\pm x+y$. The nontrivial solution of this system at
fixed $\omega\leqslant 2^{-{5}/{2}}$ corresponds to the point
where the curve of the first-order zero temperature phase
transitions and the curve of the critical points end.

\begin{figure}[!h]
\centerline{\includegraphics[width=0.55\textwidth]{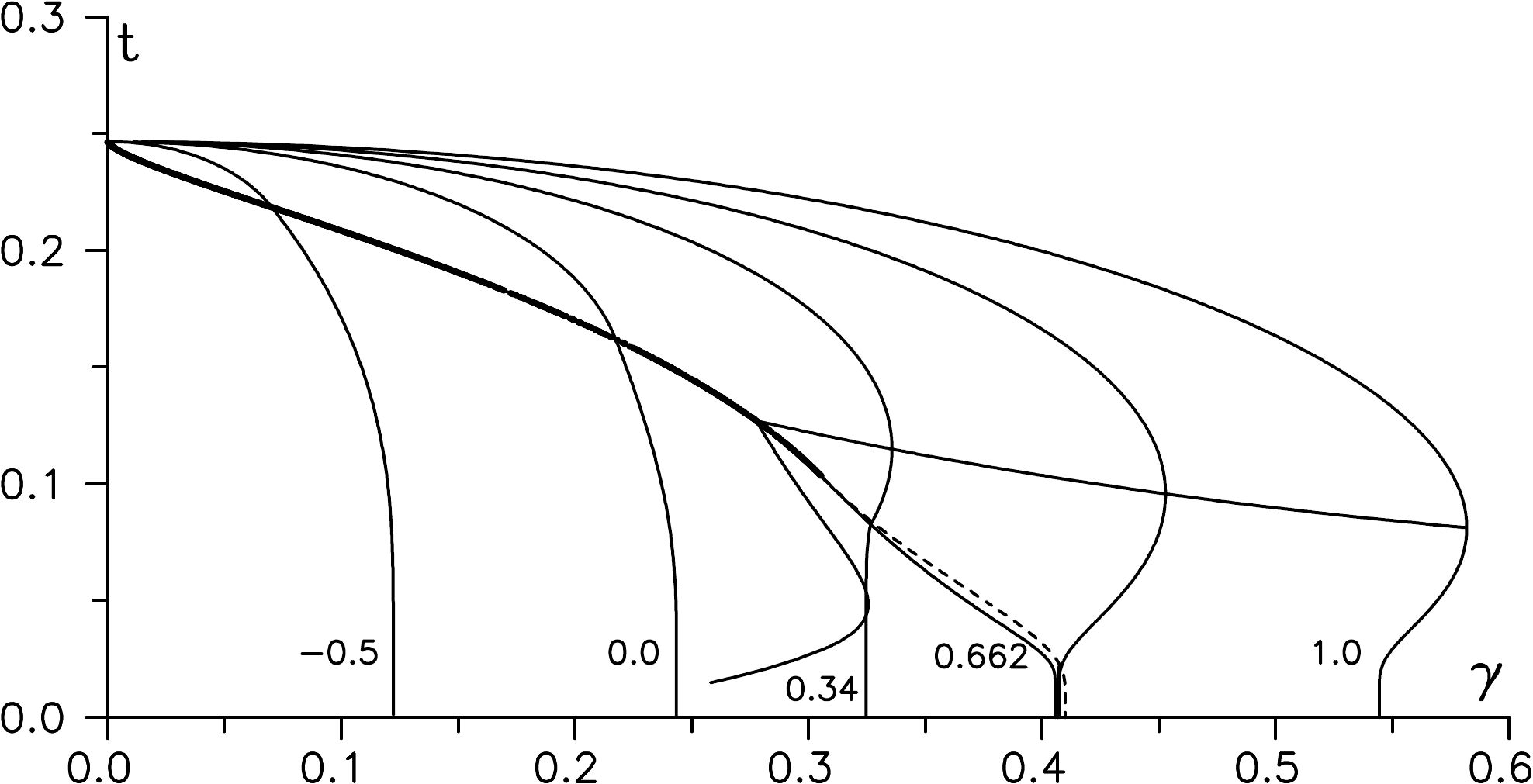}}
\caption[]{Phase coexistence curves in $(\gamma,t)$-plane for
several values of $a$ (number near curves), curve of tricritical
points (hard line), curve of extrema for the first-order phase
transitions (not hole), curve of maxima for the second-order phase
transitions, curve of branchpoints and curve of critical points
(dashed line). $\omega=0.1$.}
\label{figo1}
\end{figure}
\begin{figure}[!b]
\centerline{\includegraphics[width=0.65\textwidth]{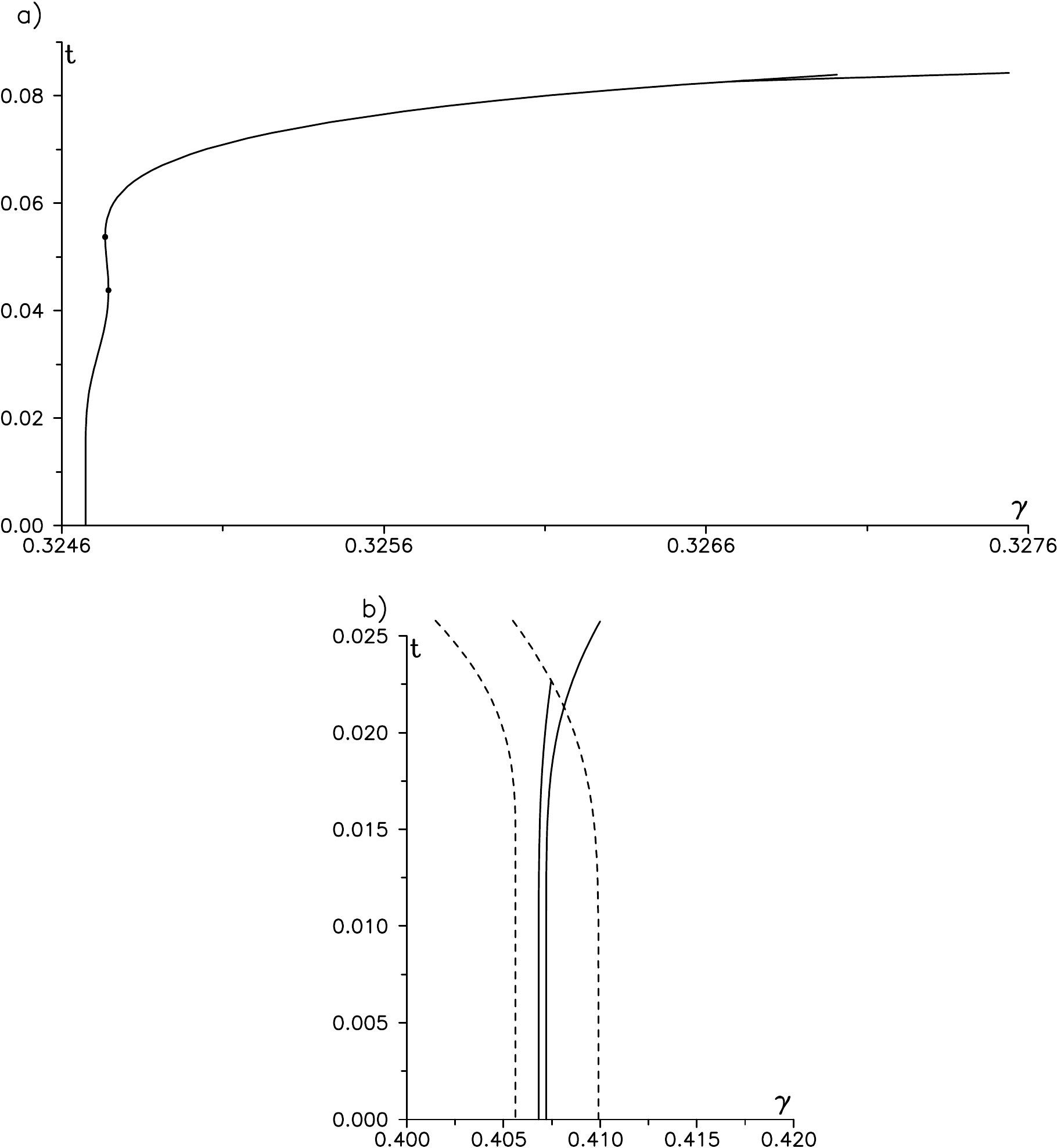}}
\caption[]{Fragments of the phase coexistence curves at
$\omega=0.1$ (see figure~\ref{figo1}). a) $a=0.34$ (circles are
the extremum points), b) $a=0.662$ (dashed line on the left is the
curve of branchpoints and the line on the right is the curve of
critical points).} \label{figo2}
\end{figure}
\begin{figure}[!h]
\centerline{\includegraphics[width=0.55\textwidth]{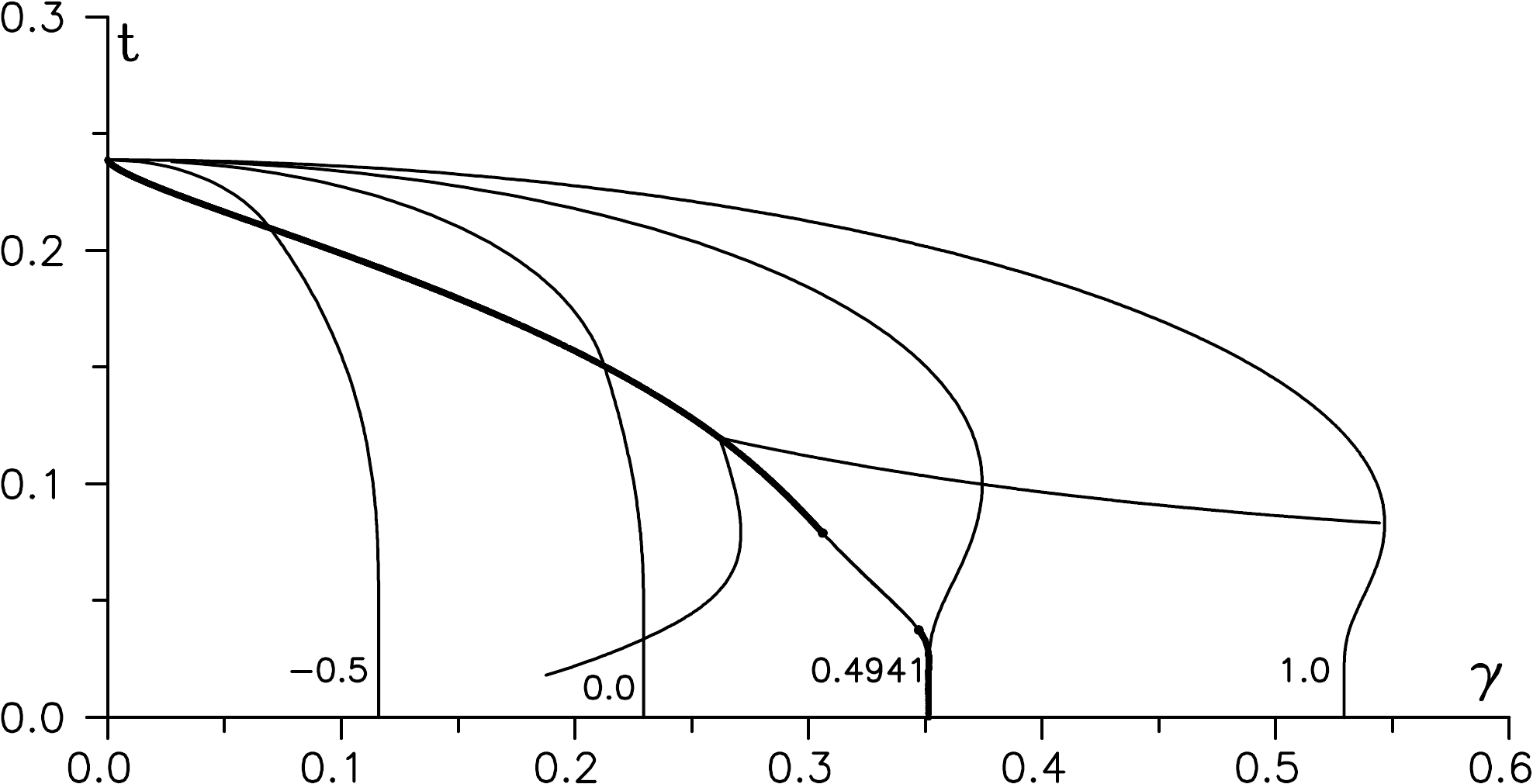}}
\caption[]{Phase coexistence curves in $(\gamma,t)$-plane for
several values of $a$ (number near curves). $\omega=0.18$. The
heavy line is the line of tricritical points. The curve of
tricritical points and the curve of branchpoints are indistinguishable
on this scale.}
\label{figo3}
\end{figure}
\begin{figure}[!h]
\centerline{\includegraphics[scale = 0.35]{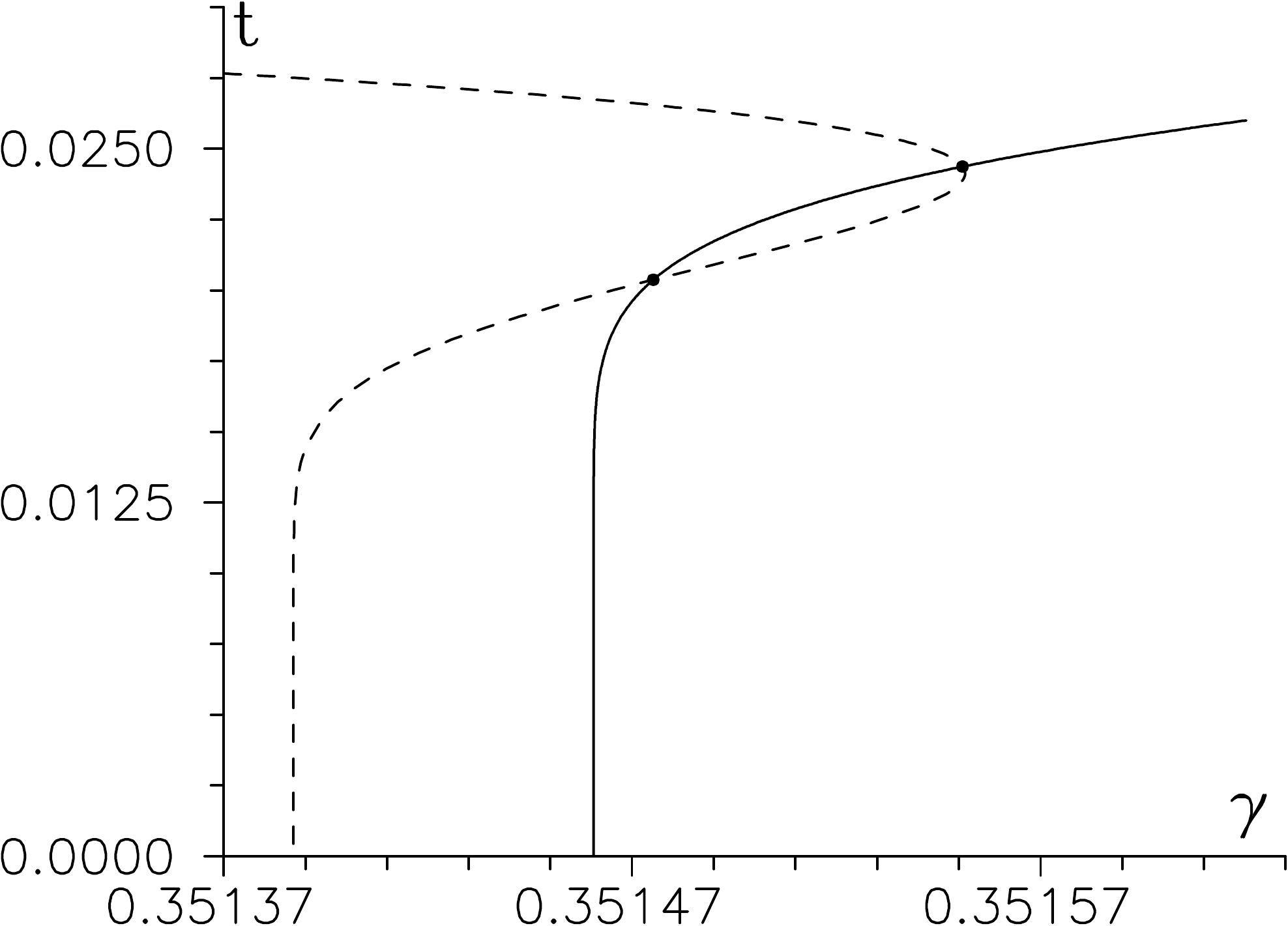}}
\caption[]{Fragment of the phase coexistence curve at $a=0.4941$,
$\omega=0.18$ (see figure~\ref{figo3}). Dashed line is the curve
of tricritical points.} \label{figo4}
\end{figure}

The phase coexistence curves in the ($\gamma, t$)-plane for
several values of $a$ as well as the curves of tricritical,
critical, and branchpoints at $\omega=0.1$ are depicted in
figure~\ref{figo1}. The latter two curves do not converge at zero
temperature and, hence, there is an interval of values of $a$
where the phase diagrams look as shown in figure~\ref{figo2}~(b),
i.e., they are composed of the curve of second-order phase
transitions and of the curve of first order phase transitions
within ferroelectric phase. At {$\omega=2^{-{5}/{2}}$} the curves
of critical points and branch- points converge again at zero
temperature, and if {$2^{-{5}/{2}}<\omega<0.196815$} these curves
converge at nonzero temperatures and pass into the curve of
tricritical points (figure~\ref{figo3}). The latter is many-valued
in low-temperature region. This leads to the existence of a new
type of phase diagram with two tricritical points
(figure~\ref{figo4}). If $\omega$ is bigger than $\approx
0.196815$, only the curve of tricritical points remains.

\subsection{Regions of existence of the ferroelectric phase}
\begin{figure}[!t]
\includegraphics[width=0.45\textwidth]{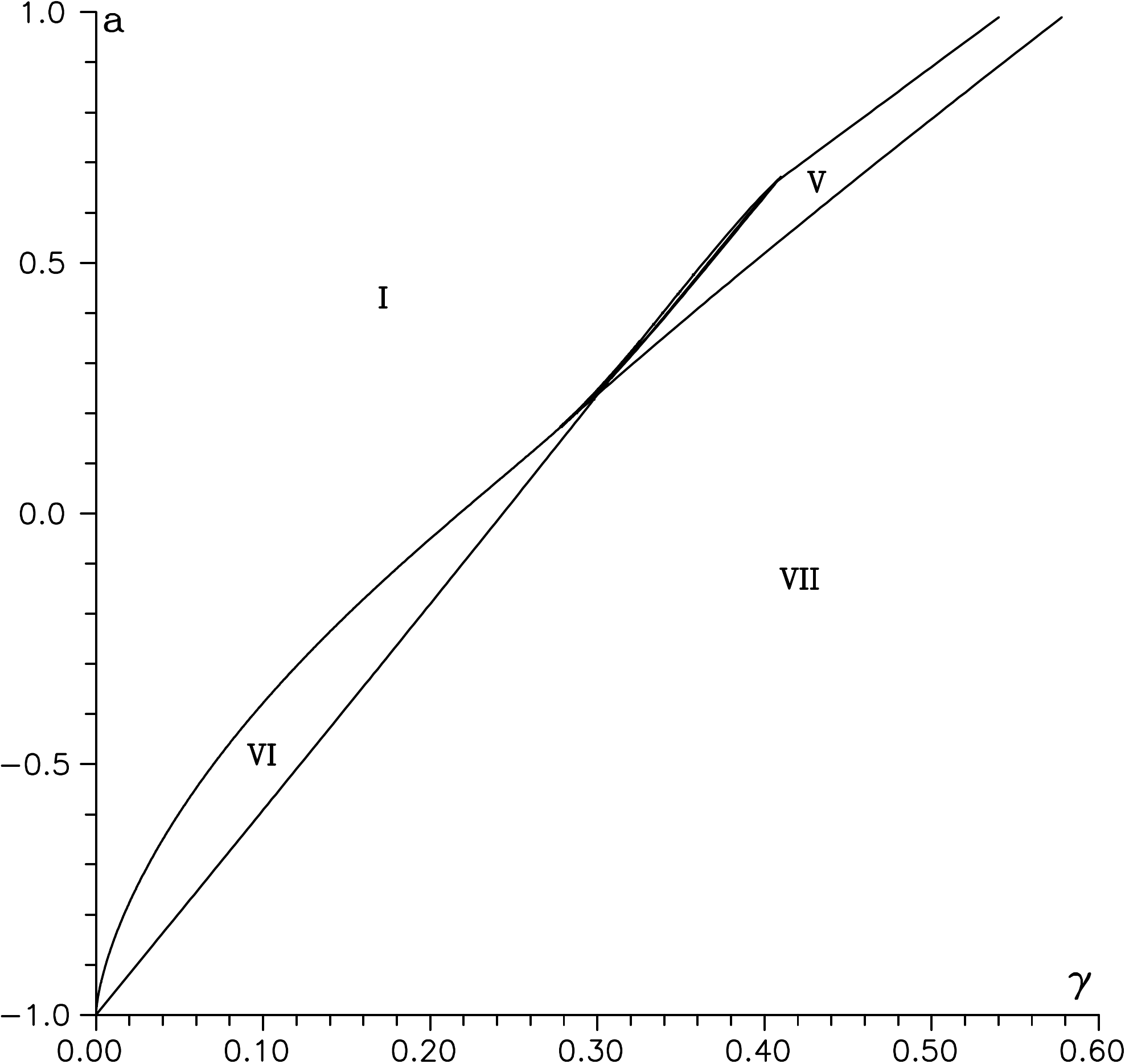}%
\hfill%
\includegraphics[width=0.47\textwidth]{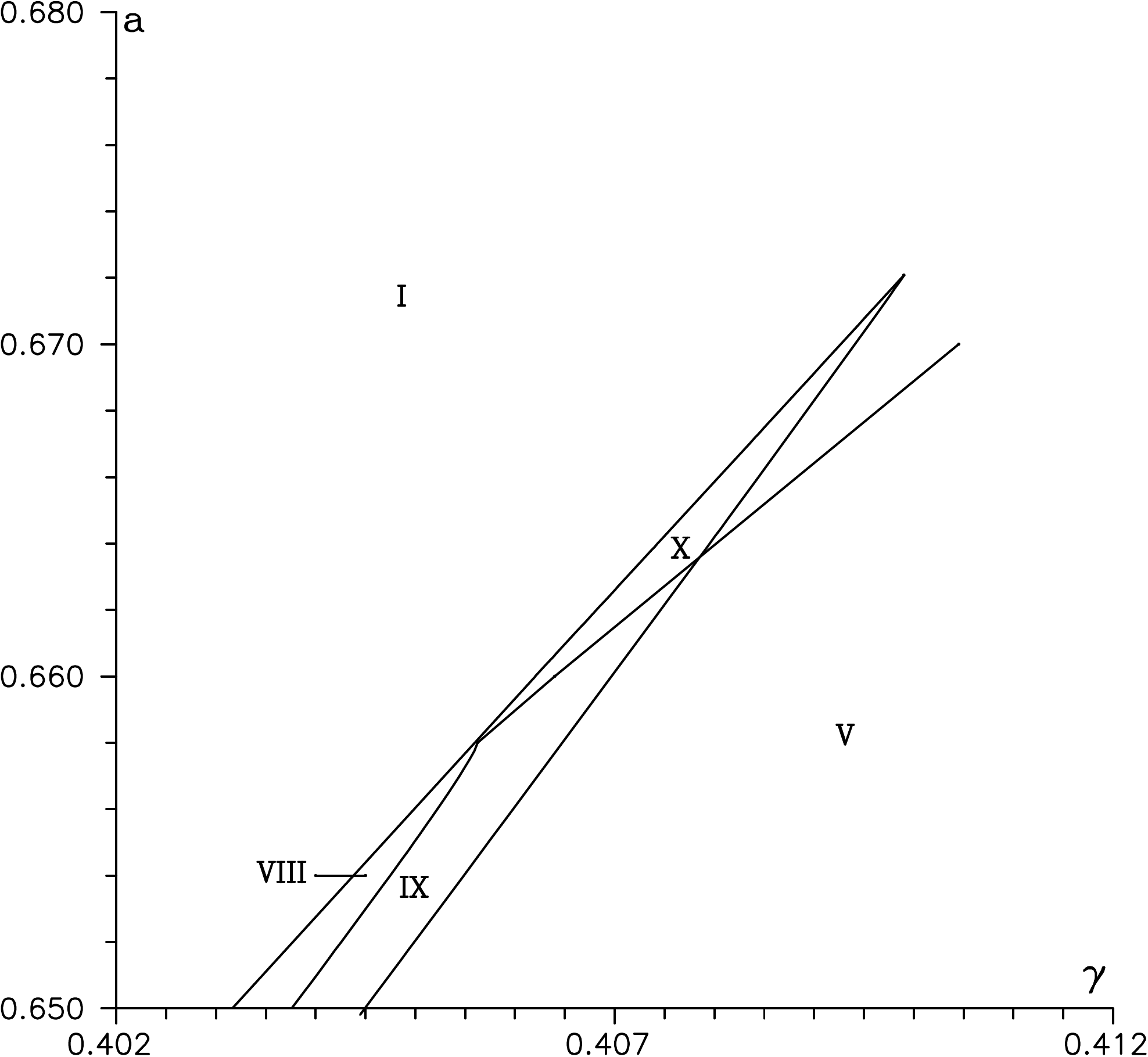}%
\\
\parbox[t]{0.49\textwidth}{%
\caption[]{Regions of existence of the ferroelectric phase.
$\omega=0.1$.}
\label{figo5}
}%
\hfill%
\parbox[t]{0.49\textwidth}{%
\caption[]{Regions of existence of the ferroelectric phase
(fragment). $\omega=0.1$. (See also figure~\ref{figo11}).}
\label{figo6}
}%
\end{figure}

\begin{figure}[!b]
\includegraphics[width=0.45\textwidth]{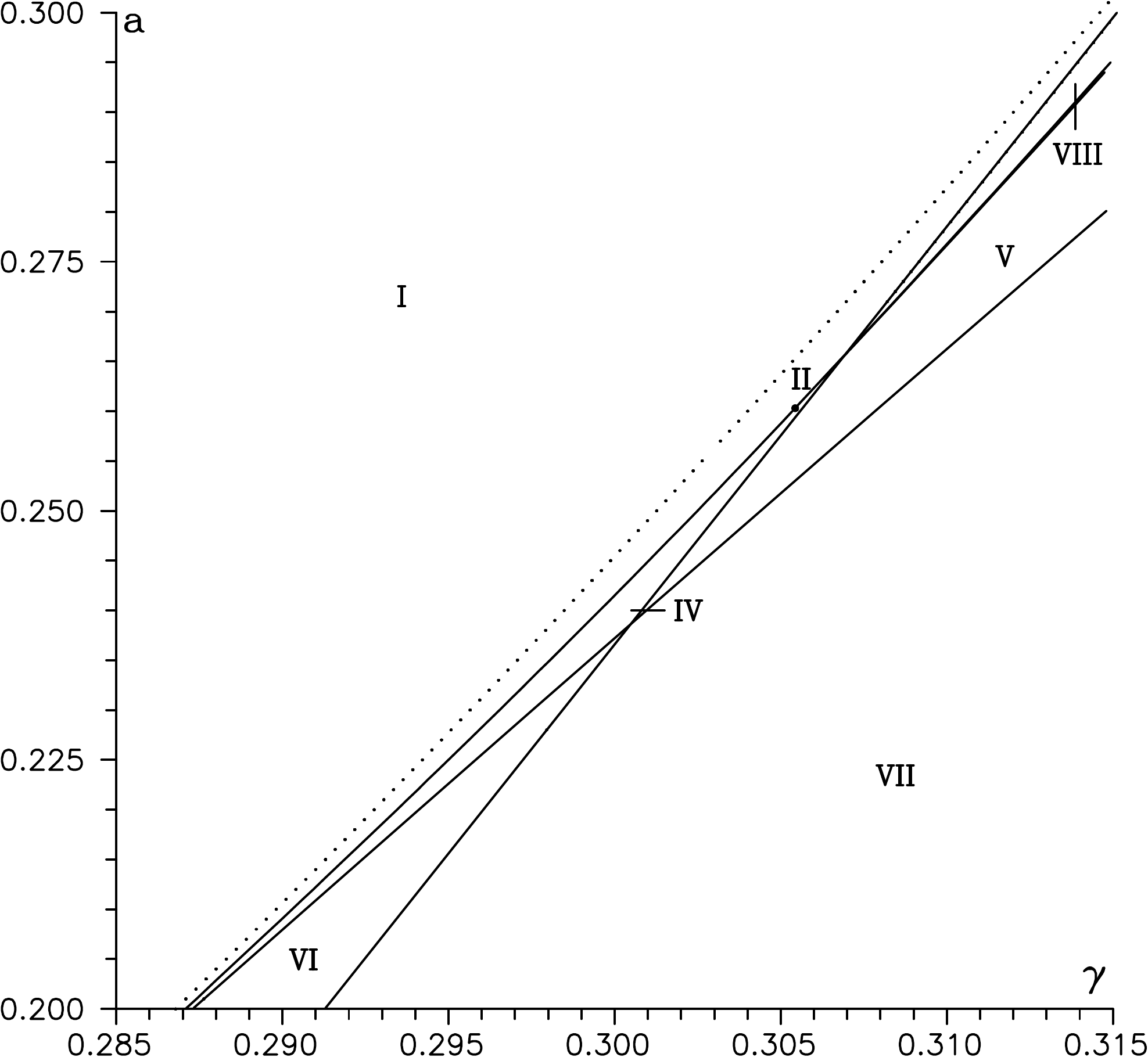}%
\hfill%
\includegraphics[width=0.47\textwidth]{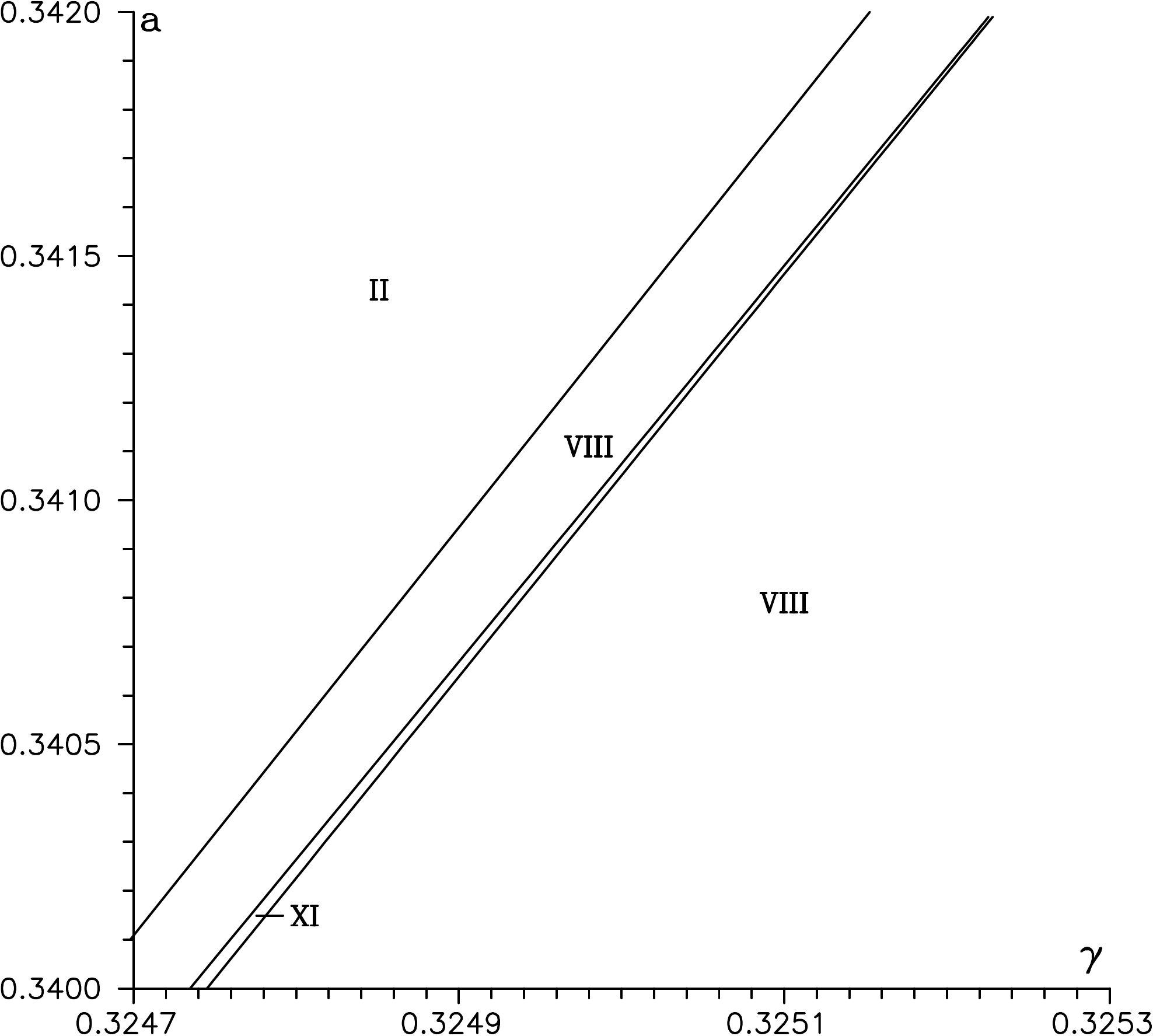}%
\\
\parbox[t]{0.49\textwidth}{%
\caption[]{Regions of existence of the ferroelectric phase
(fragment). $\omega=0.1$. The filled circle indicates the point
where the curve of tricritical points forks into the curve of
critical points and the curve of branchpoints.}
\label{figo7}
}%
\hfill%
\parbox[t]{0.49\textwidth}{%
\caption[]{Regions of existence of the ferroelectric phase
(fragment). $\omega=0.1$.}
\label{figo8}
}%
\end{figure}

\begin{figure}[!t]
\includegraphics[width=0.45\textwidth]{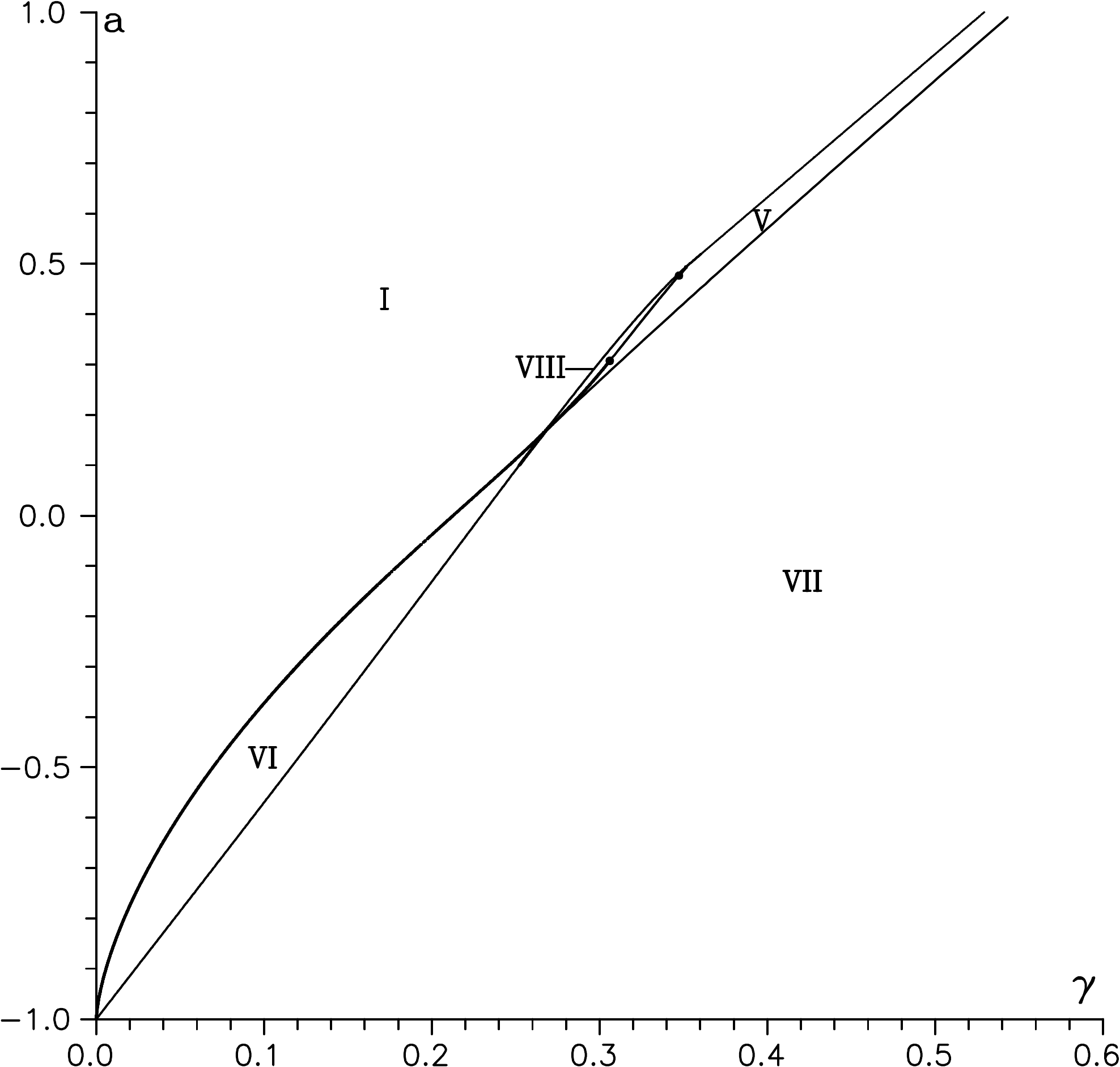}%
\hfill%
\includegraphics[width=0.49\textwidth]{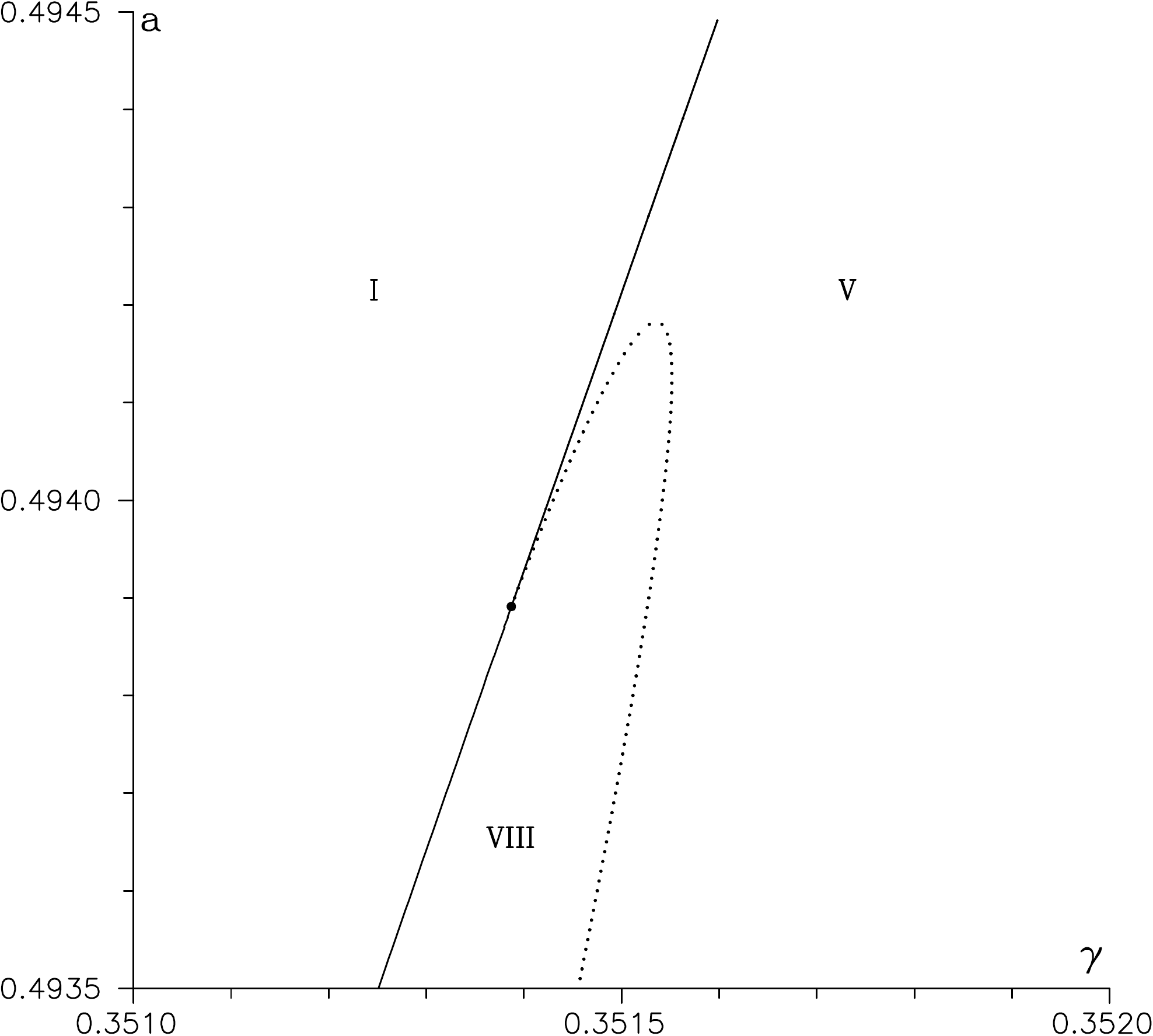}%
\\
\parbox[t]{0.49\textwidth}{%
\caption[]{Regions of existence of the ferroelectric phase.
$\omega=0.18$.}
\label{figo9}}%
\hfill%
\parbox[t]{0.49\textwidth}{%
\caption[]{Regions of existence of the ferroelectric phase
(fragment). $\omega=0.18$. Dotted line is the upper part of the
curve of tricritical points.}
\label{figo10}
}%
\end{figure}

In figures~\ref{figo5}--\ref{figo10}, the diagram of the regions
where the ferroelectric phase exists is shown for $\omega=0.1$.
The diagram is composed of the same curves as in the $\Omega = 0$
case as well as of the straight line of second-order
zero-temperature phase transitions (\ref{eqo17}) and the curve of
maxima for the first-order phase transitions (\ref{eqo20}) which
passes very closely to the curve of first-order zero-temperature
phase transitions.

%

Like in the $\omega=0$ case, two curves branch off (in the same
point) from the curve of tricritical points: the curve of minima
for the first-order phase transitions and the curve of maxima for
the second-order phase transitions. If {$\omega<0.196815$}, then
the curve of tricritical points bifurcates in the curve of
critical point and the curve of branchpoints which, if
$\omega>2^{-{5}/{2}}$, conflow again into the curve of tricritical
points. The curve of first-order zero-temperature phase
transitions is composed of two parts. The first one corresponds to
the transitions from (or into) the ferroelectric phase. It begins
at the origin of coordinates and ends at an extremity of the curve
of the branchpoints ($\omega\leqslant 2^{-{5}/{2}}$) or of the
tricritical points ($\omega>2^{-{5}/{2}}$) (this extremity is an end
point of the curve for the second-order zero-temperature phase
transitions). The second part exists only at
$\omega<2^{-{5}/{2}}$, corresponds to the phase transitions within
the ferroelectric phase and ends at an extremity of the curve of
the critical points (see figure~\ref{figo6}). In the
$\omega\geqslant 2^{-{5}/{2}}$ case, the point where the curve of
tricritical points and the curve of the first-order
zero-temperature phase transitions meet is determined by
equation~(\ref{eqo18}); in this point the latter curve smoothly
turns into the curve of the second-order zero-temperature phase
transitions.

At nonzero $\omega$ new regions occur. For instance, in region
VIII there are one low-temperature first-order phase transition
into the ferroelectric phase and one second-order phase
transition. In region IX three phase transitions exist:
first-order one and two second-order ones (the first-order phase
transition occurs within the ferroelectric phase). In region X
there are one first-order phase transition within the
ferroelectric phase and one second-order phase transition.

The curves of minima and maxima for the first-order phase
transitions converge in region VIII ($a\approx0.3433$,
$\gamma\approx0.3255$, $t\approx0.0484$) cutting off a long spit
from it (figure~\ref{figo8}). This is region XI with two
low-temperature first-order phase transitions and one second-order
phase transition.

Further, the curve of maxima for the first-order phase transitions
cuts off narrow strips from regions IX, V, and VII. These strips
are regions XII, XIII, and XIV, respectively, where, in addition,
two close low-temperature first-order phase transitions appear. In
figure~\ref{figo7} these regions are not seen because they are too
narrow.

%
%

As one can see, at a sufficiently small value of $\omega$ the
diagram is richer than at $\omega = 0$ but if the value of
$\omega$ exceeds some number, the regions of the diagram disappear
one after another and the diagram becomes poorer. Region III
disappears firstly. In figure~\ref{figo7} it looks like a short
curve segment at the beginning in the branchpoint.  It is region
XII that disappears the next, and, at $\omega=2^{-{5}/{2}}$,
region X becomes a point.

\begin{figure}[!h]
\centerline{\includegraphics[width=0.65\textwidth]{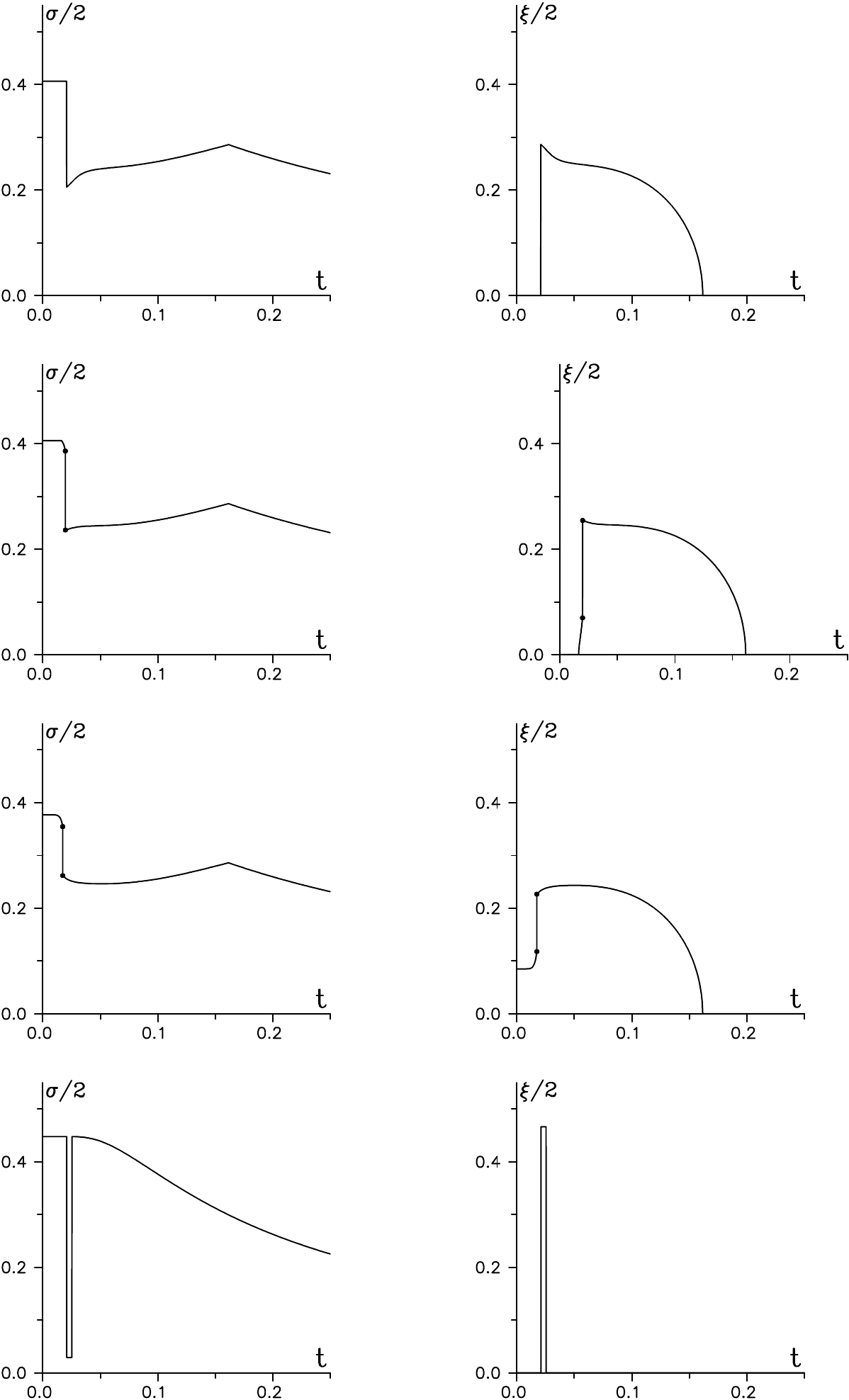}}
\caption[]{Ferroelectric and antiferroelectric order parameters
for the regions of existence of ferroelectric phase.
$\omega=0.1$. Only thermodynamically stable states are depicted.
From top to bottom: VIII) $a=0.65$, $\gamma=0.4035$; IX) $a=0.66$,
$\gamma=0.4065$; \mbox{X) $a=0.667$,} $\gamma=0.4085$; and XIV)
$a=0.2$, $\gamma=0.2912765$. The first-order phase transition
within the ferroelectric phase is shown by filled circles.}
\label{figo11}
\end{figure}

\section{Conclusions}
Hence, we performed a complete analysis of phase transitions in
the Mitsui model (without and with transverse field $\Omega$) in
the mean field approximation. Some results concerning the phase
diagram of the Mitsui model were obtained earlier~\cite{bib5,
bib12, bib13} but they were incomplete (for the $\Omega = 0$ case)
or partial (for the $\Omega \ne 0$ case).

In the $\Omega=0$ case, we derived an analytical expression for
tricritical temperature and the condition of its existence. In
this case, there are seven regions in the plane $(a,\gamma)$  that
correspond to seven different types of behavior of order
parameters. At sufficiently small but nonzero $\Omega$ their
number doubles. With $\Omega$ increasing  these regions change
their form and shift in the $(a,\gamma)$-plane. Starting with
certain value of $\Omega$ the number of regions decrease and at
$\omega\geqslant{1}/{2}$ only the region without phase transitions
remains.

At nonzero $\Omega$, second-order phase transitions are possible at
zero temperature, which is not possible at $\Omega=0$. The maximal
number of phase transitions is four and five at $\Omega = 0$ and
at $\Omega \ne 0$, respectively.

\section*{Acknowledgements}

The author is grateful to Prof. I.~Stasyuk, Prof. R.~Levitskii,
Dr.~T.~Verkholyak and Dr.~O.~Da\-ny\-liv for useful discussions.

\ukrainianpart

\title{╘рчют│ яхЁхїюфш т ьюфхы│ ╠│Ўє│}
\author{▐.▓. ─єсыхэшў}
\address{▓эёЄшЄєЄ Ї│чшъш ъюэфхэёютрэшї ёшёЄхь ═└═ ╙ъЁр┐эш,
тєы. ▓.~╤т║эЎ│Ў№ъюую, 1, 79011 ╦№т│т, ╙ъЁр┐эр}

\makeukrtitle

\begin{abstract}
\tolerance=3000%
┬ ЁюсюЄ│ т эрсышцхээ│ ёхЁхфэ№юую яюы  фюёы│фцхэю Їрчют│ яхЁхїюфш т
ьюфхы│ ╠│Ўє│ схч яючфютцэ№юую яюы , яЁюЄх ч яюяхЁхўэшь яюыхь.
┬ёЄрэютыхэю тчр║ьююфэючэрўэє чрыхцэ│ёЄ№ ь│ц Єръю■ ьюфхыы■ Єр
фтюя│ф┤ЁрЄъютю■ ьюфхыы■ Єшяє ▓ч│эур ч яючфютцэ│ь │ яюяхЁхўэшь
яюы ьш. ╧юсєфютрэю Їрчют│ ф│руЁрьш Єр ф│руЁрьш юсырёЄхщ │ёэєтрээ 
ёхуэхЄюЇрчш. ─ы  тшярфъє $\Omega = 0$ ($\Omega$~-- яюяхЁхўэх
яюых) юфхЁцрэю яЁюёЄшщ рэры│Єшўэшщ тшЁрч фы  ЄЁшъЁшЄшўэю┐
ЄхьяхЁрЄєЁш щ єьютє │ёэєтрээ  ЄЁшъЁшЄшўэю┐ Єюўъш. ─ы  $\Omega \ne
0$ чряшёрэю ёшёЄхьш Ё│тэ э№ фы  ЄЁшъЁшЄшўэю┐ Єюўъш щ єьютш ┐┐
│ёэєтрээ .

\keywords Їрчютшщ яхЁхї│ф, ёхуэхЄюЇрчр, ьюфхы№ ╠│Ўє│, ЄЁшъЁшЄшўэр
Єюўър

\end{abstract}

\end{document}